\documentclass[%
reprint,
showpacs,preprintnumbers,
aps,
prb,
]{revtex4-1}

\usepackage{graphicx}% Include figure files
\usepackage{dcolumn}% Align table columns on decimal point
\usepackage{bm}% bold math
\usepackage{amsmath}
\newcommand{\ndzo}{$\rm Nd_2Zr_2O_7$}
\begin{document}

%\preprint{APS/123-QED}

\title{Structural and Magnetic Investigations of Single-Crystals of the Neodymium Zirconate Pyrochlore, Nd$_{2}$Zr$_{2}$O$_{7}$}

\author{M. \surname{Ciomaga Hatnean}}
\homepage{go.warwick.ac.uk/supermag}
\author{M. R. Lees}
\author{O. A. Petrenko}
\author{D. S. Keeble}
\altaffiliation{Current address: Diamond Light Source, Chilton, OX11 0DE, United Kingdom}
\author{G. Balakrishnan}
\email{G.Balakrishnan@warwick.ac.uk}
\affiliation{Department of Physics, University of Warwick, Coventry, CV4 7AL, United Kingdom\\}
\author{M. J. Gutmann}
\affiliation{ISIS Facility, Rutherford Appleton Laboratory, Harwell, Oxford, Didcot, OX11 0QX, United Kingdom\\}
\author{V. V. Klekovkina}
\author{B. Z. Malkin}
\affiliation{Kazan Federal University, Kazan, 420008, Kremlevskaya 18, Russia\\}

\date{\today}% It is always \today, today,

\begin{abstract}
We report structural and magnetic properties studies of large high quality single-crystals of the frustrated magnet, \ndzo. Powder x-ray diffraction analysis confirms that \ndzo~adopts the pyrochlore structure. Room-temperature x-ray diffraction and time-of-flight neutron scattering experiments show that the crystals are stoichiometric in composition with no measurable site disorder. The temperature dependence of the magnetic susceptibility shows no magnetic ordering at temperatures down to 0.5~K. Fits to the magnetic susceptibility data using a Curie-Weiss law reveal a ferromagnetic coupling between the Nd moments. Magnetization versus field measurements show a local Ising anisotropy along the $\left\langle 111\right\rangle$ axes of the Nd$^{3+}$ ions in the ground state. Specific heat versus temperature measurements in zero applied magnetic field indicate the presence of a thermal anomaly below $T\sim7$~K, but no evidence of magnetic ordering is observed down to 0.5~K. The experimental temperature dependence of the single-crystal bulk $dc$ susceptibility and isothermal magnetization are analyzed using crystal field theory and the crystal field parameters and exchange coupling constants determined. 
\begin{description}
\item[PACS numbers] 61.50.-f, 75.10.Dg, 75.40.Cx
%61.50.-f	Structure of bulk crystals
%75.10.Dg	Crystal-field theory and spin Hamiltonians
%75.40.Cx	Static properties (order parameter, static susceptibility, heat capacities, critical exponents, etc.)

\end{description}
\end{abstract}

\pacs{Valid PACS appear here}% PACS, the Physics and Astronomy
\maketitle

%\tableofcontents

\section{Introduction} 
Pyrochlore oxides of the type $A_{2}^{3+}B_{2}^{4+}O_{7}$ (where $A=$~trivalent rare earth, $B=$~tetravalent transition metal element) are geometrically frustrated magnets~\cite{Greedan2001,Gardner2010}. These compounds have a face-centered cubic structure, with the space group $Fd\bar{3}m$ (No. 227). Both the rare earth atoms occupying the $A$ sites and the transition metal elements located on the $B$ sites form a pyrochlore lattice, a three-dimensional arrangement of corner-sharing tetrahedra, known to display the highest degree of geometrical frustration, resulting in the many unusual magnetic properties of these systems. Depending on the nature of the interaction between the magnetic ions ($A$ and/or $B$ cations), these systems can exhibit either spin ice~\cite{Harris1997,Ramirez1999}, spin glass~\cite{Gaulin1992,Gardner1999_1}, or highly correlated quantum disordered spin liquid states~\cite{Gardner1999_2}, as well as long-range magnetic order~\cite{Gardner2010}. 

One exciting avenue of research in this field has focused on systems where, due to a smaller magnetic moment on the trivalent rare earth $A$-site, quantum fluctuations play an important role in controlling the low temperature physics of the materials. Recent studies have pointed to the possibility of quantum spin liquid behavior (in which spin ice correlations exist at finite temperature, together with strong quantum effects), in a number of pyrochlore materials including Tb$_{2}$Ti$_{2}$O$_{7}$ and Yb$_{2}$Ti$_{2}$O$_{7}$~\cite{Molavian2007,Ross2011,Gingras2014}.

The success of research on pyrochlores and particularly the titanates, is due, in part, to the availability of large, high-quality single-crystals of these materials. Single-crystals of all the titanate pyrochlore family were successfully grown using the floating-zone technique~\cite{Balakrishnan1998, Gardner98, Prabhakaran11} and their magnetic ground states and properties have been investigated in great detail. (See Refs.~\onlinecite{Greedan2001,Gardner2010} and references therein.) 

As the search for geometrically frustrated magnetic pyrochlores that exhibit quantum effects widens, the research community has turned its attention to the rare-earth zirconates, $A_{2}$Zr$_{2}$O$_{7}$~\cite{otheruses}. It has been shown that the rare-earth zirconates can be stabilized into two crystallographic structures, an ordered pyrochlore phase, for the first few elements of the lanthanide series (from lanthanum to gadolinium), or a disordered fluorite phase, for the other lanthanides~\cite{Michel1974,Blanchard2012}. In addition, these oxides undergo an order-disorder phase transition, from the pyrochlore to the fluorite structure~\cite{Michel1974,Rushton2004}. The temperature at which the structural transition occurs is strongly dependent on the nature of the rare-earth element. In addition, the co-existence of both the fluorite and pyrochlore phases has been observed in some zirconates with complex compositions~\cite{Clements2011, Blanchard2013}. 

To date, due to the high melting point of the zirconates, a majority of the studies of the structural and magnetic properties of the zirconate pyrochlores have been performed on powder samples~\cite{Blote1969}. Recently, large single-crystals of the Pr containing zirconate pyrochlore, Pr$_{2}$Zr$_{2}$O$_{7}$, have been grown by the floating-zone technique~\cite{Matsuhira2009,Hatnean2014_1,Koohpayeh2014}. Studies of the magnetic properties have shown that the Pr$^{3+}$ ions in Pr$_{2}$Zr$_{2}$O$_{7}$ have a ground state doublet with a local Ising anisotropy along $\left\langle 111\right\rangle$ axes and are coupled by an antiferromagnetic exchange, but no long-range magnetic order is observed down to very low temperatures (76 mK) ~\cite{Matsuhira2009,Kimura2013_1, Kimura2013_2}. The spin ice-like correlations revealed by elastic magnetic neutron scattering, the broad excitation spectrum observed in the low-energy inelastic neutron scattering, and the reduction of the nuclear contribution to the heat capacity~\cite{Kimura2013_1} may be connected with the strongly anisotropic superexchange and multipolar interactions~\cite{Onoda2011, Lee2012} and interactions of the Pr$^{3+}$ ions with random lattice strains (strong sensitivity of the Pr$^{3+}$ ground state to lattice disorder in Pr$_{2-x}$Bi$_x$Ru$_{2}$O$_{7}$ pyrochlores was noted in Ref.~\onlinecite{Duijn14}). Some possible low temperature phases of the Pr$_{2}$Zr$_{2}$O$_{7}$ pyrochlore have been discussed recently in Ref.~\onlinecite{Gingras2014}.

%Pr$_{2}$Zr$_{2}$O$_{7}$ exhibits spin freezing at very low temperatures~\cite{Matsuhira2009,Kimura2013_1,Kimura2013_2}. Strong quantum fluctuations were observed, together with spin ice-like correlations~\cite{Kimura2013_1}. Given these recent results, the Pr$_{2}$Zr$_{2}$O$_{7}$ pyrochlore is considered to be a promising candidate for investigating the quantum spin ice phase~\cite{Kimura2013_1,Gingras2014}, although the low temperature phase of this material remains to be elucidated.

Neodymium zirconate, \ndzo, has been the subject of just a small number of studies (see for example, Ref.~\onlinecite{Blote1969}). A study of the phase diagram of Nd$_{2}$O$_{3}$-ZrO$_{2}$ shows that the pyrochlore oxide, \ndzo\ melts congruently above $2000~^{\circ}$C, although the melting point was not established~\cite{Roth1956}. The order to disorder transition in Nd$_{2}$O$_{3}$-ZrO$_{2}$ occurs at $2300~^{\circ}$C~\cite{Michel1974,Ohtani2005}, raising the possibility that it may be difficult to prepare \ndzo\ single-crystals of the pyrochlore structure due to the relatively small difference between the melting point and the structural transition temperature. When preparing this material it is therefore essential to establish the crystal structure of the \ndzo\ samples and to adjust the conditions of synthesis in order to prepare the desired phase, as has been emphasized in previous reports~\cite{Ohtani2005,Payne2013}.

We have succeeded in preparing single-crystals of the neodymium zirconate pyrochlore, \ndzo\ by the floating-zone technique. The availability of large high quality single-crystals is especially important for the study of the properties of this class of geometrically frustrated magnets and, particularly, for solving the nature of the magnetic ground state. In this paper, we report the structural characterization and investigations of the magnetic properties of single-crystals of the neodymium zirconate pyrochlore, \ndzo.

\section{Experimental details}
Single-crystals of the \ndzo\ pyrochlore oxide were prepared by the floating-zone technique, using a four-mirror Xenon arc lamp optical image furnace (CSI FZ-T-12000-X\_VI-VP, Crystal Systems Incorporated, Japan). The growths were performed in air at ambient pressure and at growth speeds in the range 10-15 mm/h. The crystal growth of \ndzo\ is described elsewhere in a more detailed paper~\cite{Hatnean2014_2}.

Room-temperature powder x-ray diffraction (XRD) measurements were performed on the as-grown boules in order to check the phase purity. Small quantities of the crystals were ground into powder and the data were then collected using a Panalytical x-ray diffractometer with Cu K$\alpha_{1}$ radiation ($\lambda$ = 1.5406 \r{A}), scanning 2$\theta$ between 10 and 110$^{\circ}$, with a step size of 0.013$^{\circ}$ in $2\theta$ and a total scanning time of 16 hours. 

In order to investigate in detail the crystal structure, single-crystal diffraction experiments were carried out on small pieces cut from the as-grown boule of \ndzo. Single-crystal x-ray diffraction measurements were performed at room temperature on an Oxford Diffraction Gemini diffractometer. A small piece of dimensions (0.28 $\times$ 0.18 $\times$ 0.13 mm$^{3}$) was cleaved from the crystal boule and affixed to a glass fiber using epoxy resin. Data over large regions of reciprocal space were collected using Mo K$\alpha$ radiation ($\lambda$ = 0.71073 \r{A}). The single-crystal x-ray diffraction data were indexed and integrated using CrysAlisPro (Agilent Technologies). The crystallographic structure was refined using ShelXL~\cite{Sheldrick2008}, as implemented in Olex2~\cite{Dolomanov2009}. 

Neutron scattering measurements were performed using the SXD beam line~\cite{Keen2006}, at the ISIS pulsed neutron source, Rutherford Appleton Laboratory. A crystal of cylindrical shape, 2.5 mm in radius and 5 mm in height, was cut from a \ndzo\ crystal boule and mounted on an Al pin with adhesive Al tape and a piece of vanadium wire. The sample was placed in a closed-cycle helium refrigerator which was then evacuated to a vacuum of $\sim 10^{-5}$~mbar. Time-of-flight neutron data were recorded at room temperature for a total of five orientations, with an exposure time of  $\sim 15$ minutes per orientation. Data reduction was accomplished using the locally available software, \textit{SXD2001}~\cite{Keen2006}. The structure was then refined using the ShelX software~\cite{Sheldrick2008}.  

Chemical composition analysis was carried out by energy dispersive x-ray spectroscopy (EDAX) using a scanning electron microscope on two pieces cut from two ends of the \ndzo\ crystal boule.

The quality of the as-grown crystals was checked and the samples were aligned for cutting using a Laue x-ray imaging system with a Photonic-Science camera system. Rectangular prism shaped samples with the [100] (tetragonal), [110] (rhombic), and [111] (trigonal) directions perpendicular to one face were cut from a \ndzo\ single-crystal boule for the magnetization measurements. The demagnetizing factors for these rectangular prisms were calculated using expressions derived by Aharoni~\cite{Aharoni1998}.

Magnetic susceptibility measurements as a function of temperature were carried out down to 0.5~K in applied magnetic fields of 1~kOe using a Quantum Design Magnetic Property Measurement System MPMS-5S SQUID magnetometer along with an i-Quantum $^{3}$He insert. Magnetization measurements were also performed as a function of magnetic field up to 70~kOe directed along specific crystallographic axes at various temperatures. Heat capacity measurements in zero applied magnetic field at temperatures from 0.5 to 400~K were carried out in a Quantum Design Physical Property Measurement System with a heat capacity option using a two-tau relaxation method.

\section{Results and Discussion}
\subsection{Single crystal growth and sample composition}
The crystal boules of the neodymium zirconate, \ndzo, were transparent to light, with a dark purple color. X-ray Laue diffraction patterns indicate that in most cases the facets of the crystals were aligned along the [111] direction. The large size and the good quality of the crystals make them suitable for most physical properties characterization measurements, including those using neutrons as a probe.

Composition analysis by EDAX of the crystals shows that the cationic ratio averages close to 1:1 for Nd:Zr over the entire length of each crystal boule. The average atomic percentages of Nd, Zr, and O were ($16.4\pm{0.3}$)\%, ($17.5\pm{0.4}$)\%, and ($66.04\pm{0.1}$)\% respectively. Given the limitations of this technique there is reasonable agreement with the expected stoichiometry.

\subsection{Crystal structure}
\begin{figure}[tb]
\begin{center}
\includegraphics[width=3.4in]{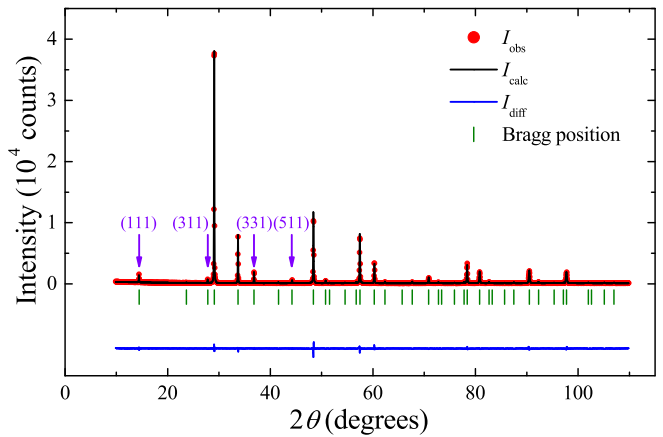}
\caption{\label{Rietveld} (Color online) Room-temperature powder x-ray diffraction pattern collected on a ground boule of \ndzo. The experimental profile (red circles) and a Rietveld refinement (black line) made using the cubic $Fd\bar{3}m$ structure are shown, with the difference given by the blue line. The positions of the Bragg peaks are indicated by green vertical bars. The positions of the (111), (311), (331), and (511) pyrochlore superlattice reflections are indicated by the violet arrows.}
\end{center}
\end{figure}

Room-temperature x-ray diffraction data were collected on small pieces of crystals that were powdered. The x-ray data were fitted to the cubic $Fd\bar{3}m$ space group~\cite{Subramanian1983}, using the Rietveld refinement method, with the Fullprof software suite~\cite{Rodriguez-Carvajal1993}. The x-ray diffraction profile for the \ndzo\ crystal displayed in Fig.~\ref{Rietveld} shows no trace of any impurity and contains a number of weak superlattice reflections demonstrating that the $Fd\bar{3}m$ pyrochlore structure is formed. These superlattice peaks, amongst which the strongest in the pattern are $hkl=(111)$, (311), (331), and (511), indicate the cation and/or anion-vacancy ordering associated with the pyrochlore structure~\cite{Blanchard2012}. Attempts to fit the x-ray data using different models for the occupancies of the Nd and the Zr sites, showed that the best model for the refinements is one in which there is no mixed occupancy of the cationic sites and no cationic/anionic deficiencies could be detected. Anisotropic atomic displacement parameters were employed in the refinement of the crystal structure. The results of the Rietveld refinement are given in Table~\ref{Powder_XRD}. The lattice parameter was found to be slightly smaller than the previously reported values for polycrystalline samples~\cite{Subramanian1983}. Analysis of the room-temperature x-ray diffraction data collected on the powder used to prepare the feed rods showed that the calculated value of the lattice parameter ($a = 10.68287(5)$~\r{A}) is, in this case, very similar to that reported in the literature~\cite{Roth1956,Subramanian1983}.

%******************Figure1*****************************

%******************Table1*****************************
\begin{table}[t]
\caption{\label{Powder_XRD}Crystallographic parameters for \ndzo\ refined using room-temperature powder x-ray diffraction data.}
\centering
\begin{tabular}{c c}
\hline
\hline
\multicolumn{2}{c}{\textbf{Structural data}} \\
\hline
Crystal system&Cubic \\
Space group&$Fd\bar{3}m$ \\
$a$ (\r{A})&10.62657(7) \\
\hline
\hline
\multicolumn{2}{c}{\textbf{Agreement factors}} \\
\hline
$R_{\mathrm{p}}$ & 7.99\% \\
$R_{\mathrm{wp}}$ & 10.5\% \\
$R_{\mathrm{exp}}$ & 6.13\% \\
$S$ & 1.71 \\
\hline
\hline
\end{tabular}
\end{table}

%******************Table2*****************************
\begin{table}[tb]
\caption{\label{Crystal_XRD}Refined structural parameters for \ndzo\ from room-temperature single-crystal x-ray diffraction data. Atomic positions for the $Fd\bar{3}m$ (origin choice 2) cubic structure are Nd, 16$d~(\frac{1}{2},\frac{1}{2},\frac{1}{2})$; Zr, 16$c~(0,0,0)$; O, 48$f~(x,\frac{1}{8},\frac{1}{8})$; and O$'$, 8$b~(\frac{3}{8},\frac{3}{8},\frac{3}{8})$. Atomic displacement parameters in units ($10^{-3}$ \r{A}$^{2}$).}
\centering
\begin{tabular}{cccccccc}
\hline
\hline
Atom&\textit{U}$_{iso}$/\textit{U}$_{eq}$&$U_{11}$&$U_{22}=U_{33}$&$U_{12}=U_{13}$&$U_{23}$\\
\hline
Nd&5.61(15)&5.61(15)&5.61(15)&-1.72(3)&-1.72(3)\\
Zr&6.4(2)&6.4(2)&6.4(2)&2.50(6)&2.50(6)\\
O&11.6(4)&15.5(10)&9.7(5)&0&2.4(8)\\
O$'$&8.5(7)&8.5(7)&8.5(7)&0&0\\
\hline
%\vspace{0.1mm}
\multicolumn{6}{l}{$a=10.62652(9)$~\r{A}, $R = 2.32\%$, $\chi^{2} = 1.224$.} \\
\multicolumn{6}{l}{O~SOF$~= 0.936(8)$, O$~x=0.3360(2)$.} \\
\hline
\hline
\end{tabular}
\end{table} 

The powder x-ray diffraction data measurements are sensitive to any preferred orientation present in the sample, and this is reflected in changes in the relative intensities of some of the Bragg peaks. In order to confirm the crystallographic parameters obtained from the powder x-ray diffraction profile, a small piece was cleaved from the crystal boule and single-crystal x-ray diffraction measurements were performed at room temperature. Attempts to fit the single-crystal x-ray data using different models for the occupancies of the cations sites showed that there is no deficiency in Nd or Zr content, and no internal site disorder was detected in our \ndzo\ crystals. The best model for the refinement of the occupancies of the anion sites was found to be one in which only the O 48$f$ site occupancy varies. Anisotropic thermal displacement parameters were used to refine the crystal structure. The structural parameters obtained from the best fit of the x-ray diffraction data at room temperature are listed in Table~\ref{Crystal_XRD}.

In order to confirm the structural model obtained using the single-crystal x-ray diffraction data, a small piece of an as-grown single-crystal of \ndzo\ was cut from the boule and single-crystal time-of-flight neutron diffraction data were collected at room temperature. Attempts to fit the neutron data using different models for the occupancies of the Nd and Zr sites, showed that the best model for the refinements is one in which the cationic occupancies are fixed to their nominal value of 1. (When the site occupancy factors (SOF) were allowed to vary, the refinement revealed an over-occupancy of the two cationic sites.) The occupancy of the oxygen $8b$ site was fixed to the value of 1, as the oxygen vacancies are usually found to be mainly on the oxygen $48f$ site~\cite{Sala2014}. (This was previously confirmed by the refinement of the single-crystal x-ray diffraction data.) The refined value of the site occupancy factor for the $48f$ site occupied by the oxygen shows that there are very few oxygen vacancies in the crystallographic structure of our \ndzo\ crystals. The crystal structure was refined using anisotropic thermal displacement parameters. A summary of the results of the Rietveld refinement of the neutron diffraction data is given in Table~\ref{Crystal_TOF}. The structural parameters obtained using the neutron diffraction data were found to be consistent with those calculated using x-ray diffraction data. 

%******************Table3*****************************
\begin{table}[tb]
\caption{\label{Crystal_TOF}Refined structural parameters for \ndzo\ from room-temperature neutron diffraction data~\cite{Refinementdata}. Atomic positions for the $Fd\bar{3}m$ (origin choice 2) are the same as in Table~\ref{Crystal_XRD}. Atomic displacement parameters in units ($10^{-3}$ \r{A}$^{2}$).}
\centering
%\hline
%Atom&Wyckoff&$x/a$&$y/b$&$z/c$&SOF&\textit{U}$_{iso}$/\textit{U}$_{eq}$ \\
% &position& & & & & \\
%\hline
%Nd&16\textit{d}& 0.5&0.5&0.5&1&5.6(5) \\
%Zr&16\textit{c}&0&0&0&1&9.3(6) \\
%O&48\textit{f}&0.3360(3)&0.125&0.125&0.97(1)&11.5(6) \\
%O$'$&8\textit{b}&0.375&0.375&0.375&1&6.9(9) \\
\begin{tabular}{cccccccc}
\hline
\hline
Atom&\textit{U}$_{iso}$/\textit{U}$_{eq}$&$U_{11}$&$U_{22}=U_{33}$&$U_{12}=U_{13}$&$U_{23}$\\
\hline
Nd&5.6(5)&5.6(5)&5.6(5)&-2.0(5)&-2.0(5)\\
Zr&9.3(6)&9.3(6)&9.3(6)&2.8(7)&2.8(7)\\
O&11.5(6)&15.5(12)&9.6(7)&0&3.1(9)\\
O$'$&6.9(9)&6.9(9)&6.9(9)&0&0\\
\hline
\multicolumn{6}{l}{ $a = 10.618(3)$~\r{A}, $R = 6.93\%$, $\chi^{2} = 1.627$.} \\
\multicolumn{6}{l}{O~SOF$~= 0.97(1)$, O$~x=0.3360(3)$.} \\
\hline
\hline
\end{tabular}
\end{table} 

%******************Figure2*****************************
%\begin{figure}[ht]
%\begin{center}
%\includegraphics[width=3.4in]{FIG2}
%\caption{\label{Structure} (Color online) Representation of the pyrochlore lattice viewed along (a) the [010] and (b) the [111] direction. The atomic sites are indicated, as follows: Nd ($16d$) in blue Zr ($16c$) in violet, O ($48f$) in yellow and O$'$ ($8b$) in orange.}
%\end{center}
%\end{figure}

The value of $x=0.3360(3)$ for the O atom in the $48f$ site confirms that both the polyhedra formed around the $A$ (Nd) and $B$ (Zr) sites are distorted from the ideal geometries (a perfect cube and a perfect octahedron respectively)~\cite{Subramanian1983}. This fact is strongly reflected in the uniaxial nature of the $A$ site symmetry and in the magnetic properties of the system. The scalenohedron NdO$_{8}$ is formed by two short bonds Nd-O$'$ of 2.2988(2)~\r{A} length, and six bonds Nd-O of 2.561(2)~\r{A}, while the trigonal antiprism ZrO$_{6}$ has all six bonds Zr-O of 2.087(1)~\AA. The two short bonds Nd-O$'$ are stacked along the [111] direction, the other six long bonds (Nd-O) connect the Nd ion at the center with oxygen (O) ions at the apexes of the trigonal antiprism strongly compressed along the [111] axis.

The structural features of the \ndzo\ crystals are very similar to those of the neighboring praseodymium zirconate, Pr$_{2}$Zr$_{2}$O$_{7}$. Rietveld refinement of the neutron diffraction data collected at the SXD beam line on a Pr$_{2}$Zr$_{2}$O$_{7}$ single-crystal indicates a value of $x=0.3349(1)$ for the O atom in the $48f$ site~\cite{Hatnean2014_3}. The oxygen polyhedra formed around the Pr and the Zr ions show a similar degree of distortion to the those of neodymium zirconate. The bond lengths inside the two coordination polyhedra are as follows: 2.3088(2)~\r{A} for the Pr-O$'$ bonds, 2.578(1)~\r{A} for Pr-O bonds, and 2.0917(8)~\r{A} for the Zr-O bonds. Given the similarities between the structural parameters of the \ndzo\ and Pr$_{2}$Zr$_{2}$O$_{7}$ pyrochlores, we can expect some similarities between the crystal fields in these two systems.

In summary, Rietveld refinements of the powder and single-crystal x-ray diffraction and neutron diffraction data confirm that \ndzo\ adopts the cubic $Fd\bar{3}m$ pyrochlore structure. The results show no measurable cationic or anionic deficiencies of the as-grown crystals of \ndzo\ and confirm the good quality of the boules and their suitability for future investigations, including those using neutrons as a probe, given the large size of the crystal boules grown by the floating-zone technique.

\subsection{\label{MagHC}Magnetization and heat capacity}
\begin{figure*}[tb]
\begin{center}
\includegraphics[width=6.8in]{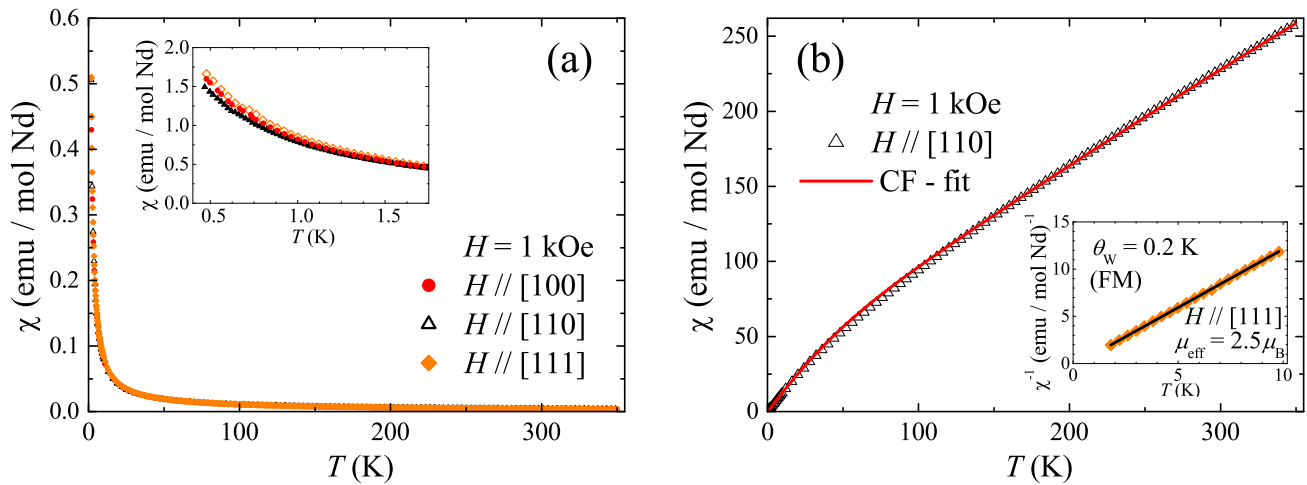}
\caption{\label{Susceptibility} (Color online) (a) Temperature dependence of the \textit{dc} magnetic susceptibility, $\chi$ versus $T$, in the temperature range 1.8 to 350~K for a crystal of \ndzo, with a magnetic field applied along the [100] (red), [110] (black), and [111] (orange) directions. The inset shows $\chi$ versus $T$ in the temperature range 0.5 to 1.75~K. (b) Measured (symbols) and calculated (the solid curve) temperature dependence of the reciprocal of the bulk $dc$ susceptibility, $\chi^{-1}$ versus $T$, of \ndzo\ for a field applied along the [110] direction. The temperature dependence of the reciprocal bulk $dc$ susceptibility was calculated using the crystal field parameters determined in the present work. The inset shows $\chi^{-1}$ versus $T$ and the linear fit (using the Curie-Weiss law) to the data in the temperature range 1.8 to 10~K for a magnetic field applied along the [111] direction.}
\end{center}
\end{figure*}

Field-cooled (FC) and zero-field-cooled (ZFC) magnetization versus temperature curves were measured on a \ndzo\ single-crystal aligned along three different crystallographic directions ([100], [110], and [111]). The temperature dependence of the \textit{dc} magnetic susceptibility,  $\chi\left(T\right)$, and reciprocal \textit{dc} magnetic susceptibility $\chi^{-1}\left(T\right)$ are shown in Fig.~\ref{Susceptibility}. No anomalies in $\chi(T)$ are observed down to 0.5~K suggesting the absence of a magnetic transition. The magnetic susceptibility data measured along the different directions all exhibit a monotonic decrease when warming from 0.5 to 350~K, and at temperatures above 5~K the magnetic susceptibilities collected in 1~kOe along the three high symmetry directions all overlap to within experimental error.

%******************Figure2*****************************

Attempts to fit the temperature dependent reciprocal magnetic susceptibilities reveal that the $\chi^{-1}\left(T\right)$ data do not obey a Curie-Weiss law in the temperature range 0.5 to 350~K. Nevertheless, we have made fits to a Curie-Weiss law over a reduced temperature range (1.8 to 10~K) (see Fig.~\ref{Susceptibility}b inset), and the results of these fits have shown that for $T<10$~K \ndzo\ has an effective moment of $\mu_\mathrm{{eff}}=2.543(2)\mu_\mathrm{{B}}$ ($\mu_\mathrm{{B}}$ is the Bohr magneton) and a Weiss temperature of $\theta_\mathrm{{W}}=+0.200(8)$~K. Over an extended temperature range from 1.8 to 60~K, the measured temperature dependence of the $dc$ susceptibility can be approximated by the expression $\chi\left(T\right)=C/\left(T-\theta_{\mathrm{W}}\right)+\chi_{\mathrm{VV}}$ with the Weiss temperature of $\theta_{\mathrm{W}}=+0.150$~K and a Van Vleck contribution of $\chi_{\mathrm{VV}}=0.00346$~emu/(mol Nd). The effective moment of Nd$^{3+}$ in \ndzo\ is smaller than the magnetic moment of $\frac{12}{\sqrt{11}}\mu_{\mathrm{B}}$ of a free Nd$^{3+}$ in the ground state $^4I_{9/2}$. The $\theta_\mathrm{{W}}$ value indicates a ferromagnetic coupling between the Nd spins. We have also performed magnetic susceptibility measurements on a ground piece of the \ndzo\ crystal boule. The estimated Weiss temperature $\theta_\mathrm{{W}}$ was, in this case, found to be +0.108(6)~K. Previous studies also reported small absolute values for the Weiss temperature in Nd-based pyrochlores, with $\theta_\mathrm{{W}}=+0.06$~K in \ndzo~\cite{Blote1969}, $\theta_\mathrm{{W}}=-0.069(4)$~K in Nd$_{2}$Pb$_{2}$O$_{7}$~\cite{Hallas15}, and $\theta_\mathrm{{W}}=-0.31$~K in Nd$_{2}$Sn$_{2}$O$_{7}$~\cite{BondahJagalu01, Matsuhira2002}. Small differences in the absolute values obtained for \ndzo\ may be explained by the fact that some measurements were carried out on aligned crystal, along well-defined crystallographic axes, while others were for polycrystalline samples. The results also appear to depend on the exact temperature range over which the fits are performed. The difference between the data collected for different directions of the applied field at low temperatures (see inset in Fig.~\ref{Susceptibility}a) cannot be explained by the contribution of the demagnetizing field; the difference in the susceptibilities corrected for the demagnetization factor has been found to be close to 1\%, for all the crystallographic directions. However, this difference, which is found in a small but finite magnetic field of 1~kOe, and which increases with decreasing temperature, may signal the onset of a phase transition at lower temperature. Different signs of $\theta_\mathrm{{W}}$ in compounds with different $B$-site cations (a chemical pressure effect) may be caused by closely competing exchange and dipolar interactions between the Nd$^{3+}$ ions. 

\begin{figure}[tb]
\begin{center}
\includegraphics[width=3.4in]{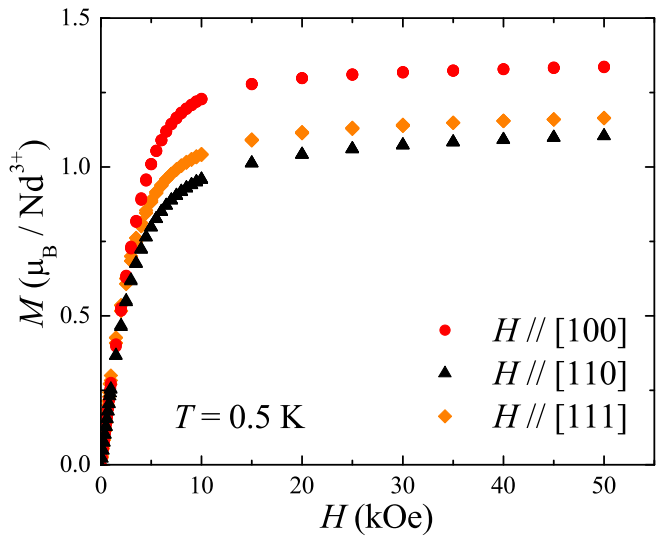}
\caption{\label{Magnetization} (Color online) Isothermal magnetization ($M$) as a function of applied magnetic field ($H$) along the [100] (red), [110] (black), and [111] (orange) directions at $T=0.5$~K for a single-crystal of \ndzo.}
\end{center}
\end{figure}

\begin{figure}[tb]
\begin{center}
\includegraphics[width=3.4in]{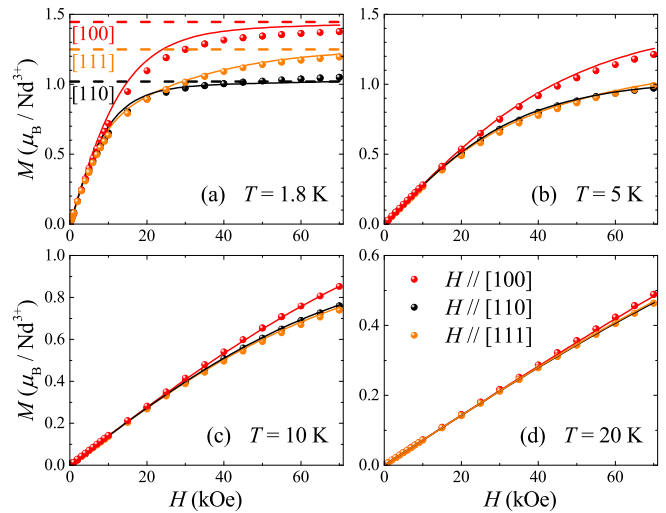}
\caption{\label{Magn_temp} (Color online) Measured (symbols) and calculated (solid lines) external magnetic field dependence of the magnetization ($M$) of \ndzo\ at various temperatures. The field was applied along the [100] (red), [110] (black), and [111] (orange) axes of the \ndzo\ crystal lattice, at temperatures of (a) 1.8~K, (b) 5~K, (c) 10~K , and (d) 20~K. The field dependence of the magnetization was calculated using the crystal field parameters determined in the present work. The dashed lines in (a) indicate the theoretically expected values of the saturated magnetization satisfying the ratios $\sqrt{2}$:$\sqrt{3/2}$:1 for the [100], [111], and [110] directions, respectively, in a classical spin ice configuration~\cite{Fukazawa2002}.}
\end{center}
\end{figure}

Figure~\ref{Magnetization} shows the magnetization as a function of applied magnetic field $M\left(H\right)$ at a temperature of $T=0.5$~K for \ndzo\ along three high symmetry directions, [100], [110], and [111]. The magnetization response is reversible with no hysteresis between the field increasing and field decreasing $M\left(H\right)$ curves. The data reveal a non-linear variation of the magnetization as a function of applied field. The magnetization is highest in strong magnetic fields applied along the [100] direction and lowest for the [110] direction. The field dependence of the magnetization along the three high symmetry directions, measured at various temperatures (see Fig.~\ref{Magn_temp}) indicate that the magnetic anisotropy evolves on cooling. The results of the magnetization measurements suggest that the magnetization follows a local Ising behavior, with a different dependence of $M\left(H\right)$ along different crystallographic directions. This kind of magnetic anisotropy represents a key feature of the spin ice system, as observed in the Dy$_{2}$Ti$_{2}$O$_{7}$ and Ho$_{2}$Ti$_{2}$O$_{7}$ pyrochlore oxides~\cite{Fukazawa2002,Petrenko2003}. However, the field dependent magnetization data reveal that the magnetic anisotropy observed for the \ndzo\ pyrochlore is only partially consistent with the spin ice model, as the values of the magnetic moments measured for the [100], [110], and [111] directions, at the maximum applied field do not agree exactly with the ratios of the expected saturated moments for a classic spin ice configuration~\cite{Fukazawa2002} (see Fig.~\ref{Magn_temp}a) and that in contrast to the titanate spin ice materials, there is no evidence for a shoulder developing in $M\left(H\right)$ when $H$ is applied along [111]. These results are, however, similar to those reported for the related system Pr$_{2}$Zr$_{2}$O$_{7}$~\cite{Kimura2013_1,Kimura2013_2,Hatnean2014_1} and provide evidence for a mixing of the wave functions of the ground-state doublet and excited crystal field (CF) levels of the Nd$^{3+}$ and Pr$^{3+}$ ions in strong external magnetic fields.

%******************Figure3*****************************

%******************Figure4*****************************

\begin{figure}[tb]
\begin{center}
\includegraphics[width=3.4in]{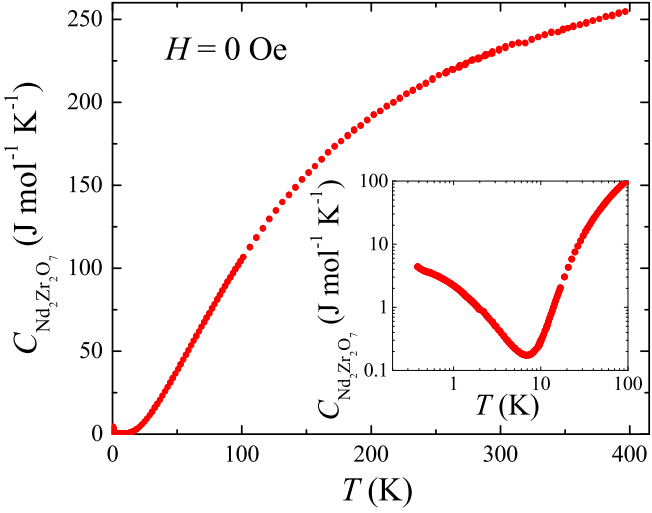}
\caption{\label{Exp_heat_capacity} (Color online) Heat capacity as a function of temperature in zero applied magnetic field for a \ndzo\ single crystal. The inset shows the temperature dependence of the heat capacity data on a logarithmic scale.}
\end{center}
\end{figure}

Specific heat measurements were performed in the temperature range 0.5 to 400~K. The temperature dependence of the heat capacity of \ndzo\ in zero external magnetic field is shown in Fig.~\ref{Exp_heat_capacity}. The data agree well with the previously published results for \ndzo~\cite{Blote1969,Lutique2003_2}. A continuous decrease in the specific heat with decreasing temperature is followed by an upturn at temperatures below $\sim7$~K. Previous heat capacity studies performed on a polycrystalline sample of \ndzo\ show a phase transition centered at 0.37~K and evidence of some magnetic ordering~\cite{Blote1969}. 

%Our current measurements, however, do not extend below 0.5~K.

%******************Figure5*****************************

\subsection{Crystal field parameters}
To analyze the magnetic behavior of \ndzo, we consider the following Hamiltonian of a single Nd$^{3+}$ ion in an applied magnetic field $H$:

\begin{equation} \label{eq1}
h = h_{\mathrm{FI}} + h_{\mathrm{CF}} + h_{\mathrm{Z}}.
\end{equation}

\noindent Here $h_{\mathrm{FI}}$ is a free ion standard Hamiltonian~\cite{Crosswhite1984} that operates in the total space of 364 states of the electronic $4f^{3}$ configuration. The next term in (\ref{eq1}),

\begin{eqnarray} \label{eq2}
h_{\mathrm{CF}} = B_{2}^{0}C_{0}^{2} + B_{4}^{0}C_{0}^{4} + B_{4}^{3}(C_{3}^{4} - C_{-3}^{4}) + B_{6}^{0}C_{0}^{6} \nonumber 
\\ + B_{6}^{3}(C_{3}^{6} - C_{-3}^{6}) + B_{6}^{6}(C_{6}^{6} + C_{-6}^{6}),
\end{eqnarray}

\noindent where $h_{\mathrm{CF}}$ is the energy of 4$f$ electrons in the crystal field of the perfect lattice with the $D_{3d}$ symmetry at the Nd$^{3+}$ sites written in the local system of coordinates with the $z$ axis along the trigonal symmetry axis. $B_{p}^{q}$ are CF parameters and $C_{q}^{p}$ are spherical tensor operators. The third term in (\ref{eq1}) is,

\begin{equation} \label{eq3}
h_{\mathrm{Z}} = -\boldsymbol{\mu}\cdot \bm{H}_{\mathrm{loc}}.
\end{equation}

\noindent where $h_{Z}$ is the electronic Zeeman energy. $\boldsymbol{\mu} = -\mu_\mathrm{{B}}(k\bm{L}+2\bm{S})$ is the magnetic moment of a Nd$^{3+}$ ion ($\bm{L}$ and $\bm{S}$ are the electronic orbital and spin moments, respectively, and $k$ is the orbital reduction factor~\cite{Malkin2010}). The local magnetic field affecting the Nd$^{3+}$ ions, $\bm{H}_{\mathrm{loc}} = \bm{H} + \bm{H}_{\mathrm{exch}} + \bm{H}_{\mathrm{dip}} - \bm{H}_{\mathrm{D}}$, involves the external field, $ \bm{H}$, the exchange and dipolar fields, $\bm{H}_{\mathrm{exch}}$ and $\bm{H}_{\mathrm{dip}}$, respectively, corresponding to anisotropic exchange and magnetic dipolar interactions between the Nd$^{3+}$ ions that are considered within the self-consistent field approximation , and the demagnetizing field $\bm{H}_{\mathrm{D}} = 4\pi N\bm{M}/3$, where $N$ is the demagnetizing factor, $\bm{M} = \sum_{n} \bm{m}_{n}/\nu$ is the magnetization, and $\nu=a^3/4$ is the unit cell volume. There are four magnetically non-equivalent rare-earth ions in the unit cell and $\bm{m}_{n}$ is the average magnetic moment of a Nd$^{3+}$ ion belonging to the sublattice $n$.

The dipolar fields at sites $n$, $H_{\mathrm{dip},n\alpha} = \sum_{n'\beta}\rho Q_{n\alpha}^{n'\beta} m_{n'\beta}$ ($\rho = 4\pi/3\nu$), are determined by dimensionless lattice sums $Q_{n\alpha}^{n'\beta}$ which have been computed in Ref.~\onlinecite{Malkin2010}. The exchange interaction is assumed to be non-zero for the nearest neighbor Nd$^{3+}$ ions only. In particular, the exchange interaction between the ions with radius vectors $\bm{r}_{1}=\left(\frac{1}{8}, \frac{1}{8}, \frac{1}{8}\right)a$ and $\bm{r}_{2}=\left(-\frac{1}{8}, -\frac{1}{8}, \frac{1}{8}\right)a$ is approximated by:~\cite{Malkin2010}

\begin{eqnarray} \label{eq4}
h_{\mathrm{exch}} (1,2) = -\lambda_{\parallel} \mu_{1x'} \mu_{2x'} - \lambda_{\perp 1} \mu_{1z'} \mu_{2z'} \nonumber
\\ - \lambda_{\perp 2} \mu_{1y'} \mu_{2y'},
\end{eqnarray}

\noindent where $\mu_{x'}$, $\mu_{y'}$, and $\mu_{z'}$ are operators of the magnetic moment components along the vectors $\bm{X}' = \bm{r}_{1} - \bm{r}_{2}$, $\bm{Y}' = \bm{r}_{2} \times \bm{r}_{1}$, and $\bm{Z}' = \bm{r}_{1} + \bm{r}_{2}$. To reduce the number of unknown parameters, we neglect in~(\ref{eq4}) an additional antisymmetric term $\lambda_{\mathrm{DM}}\bm{Y}'\cdot\left(\boldsymbol{\mu}_1\times\boldsymbol{\mu}_2\right)$ (Dzyaloshinskii-Moriya like) which is allowed by symmetry of the pyrochlore lattice~\cite{Petit2014}. The values of the exchange coupling constants $\lambda_{\parallel}$, $\lambda_{\perp 1}$, and $\lambda_{\perp 2}$,  six CF parameters, and the orbital reduction factor were obtained from the fitting procedure by comparing the measured and computed magnetic field dependence of the heat capacity, the bulk $dc$ susceptibility with temperature, and the magnetic field dependence of the isothermal magnetization. 

\begin{figure}[tb]
\begin{center}
\includegraphics[width=3.4in]{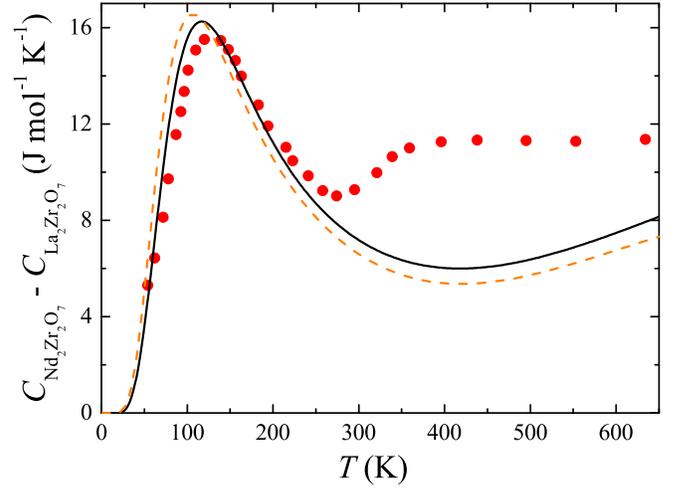}
\caption{\label{Heat_capacity} Calculated and experimentally estimated contributions to the heat capacity of \ndzo, due to CF excitations in the Nd$^{3+}$ ions ions (the temperature range is extended up to 650~K using the data taken from Ref.~\onlinecite{Sedmidubsky2005}). The solid and dashed curves correspond to the results of calculations based on the CF energies obtained in the present work and in Ref.~\onlinecite{Lutique2004}, respectively.}
\end{center}
\end{figure}

Modeling of the magnetization, $dc$ susceptibility, and the contribution of the Nd subsystem to the heat capacity involved numerical diagonalization of the Hamiltonian~(\ref{eq1}) for fixed values of the applied magnetic field and subsequent quantum-statistical averaging of the single-ion energy, $\textless h\textgreater$, the square of energy, $\textless h^2\textgreater$, and the magnetic moment components for different temperatures. Parameters of the free ion Hamiltonian, $h_{\mathrm{FI}}$, were taken from Ref.~\onlinecite{Popova2007}. The initial values of the CF parameters were calculated in the framework of the exchange charge model~\cite{Klekovkina2014, Malkin2004} by making use of the parameters of the crystal lattice determined in the present work (see column A in Table~\ref{CFparameters}). The effective ionic charges were taken from Ref.~\onlinecite{Malkin2004} and the scaling factors which determine the exchange charges on the Nd-O$'$ and Nd-O bonds were obtained from the analysis of the contribution to the specific heat from the CF excitations in the Nd subsystem. This contribution as a function of temperature was estimated from the difference between the measured heat capacities of \ndzo\ and non-magnetic La$_2$Zr$_2$O$_7$ given in Ref.~\onlinecite{Sedmidubsky2005}. The results of calculations of the heat capacity of the Nd~subsystem $C_{\mathrm{Nd}}=N_{\mathrm{A}}\left(\textless h^2\textgreater - \textless h\textgreater^2\right)/k_\mathrm{B}T^2$ ($N_\mathrm{A}$ is the Avogadro number, $k_\mathrm{B}$ is the Boltzmann constant) match well the experimental data for temperatures below 300~K (see Fig.~\ref{Heat_capacity}). More detailed analysis, in particular taking into account changes of phonon frequencies~\cite{Lan15}, is necessary to clarify the behavior of the relatively small differences between the heat capacities of neodymium and lanthanum zirconates at higher temperatures.

%The corresponding scaling factors in calculations of the crystal field parameters were estimated by taking into account the contribution to the specific heat from the crystal field excitations in the Nd subsystem as presented in Ref.~\onlinecite{Sedmidubsky2005}. The heat capacity of \ndzo\ was presented in Refs.~\onlinecite{Lutique2003_1} and \onlinecite{Lutique2003_2}, and the differences between the heat capacities of \ndzo\ and La$_{2}$Zr$_{2}$O$_{7}$ are well reproduced by our calculations for temperatures below 300 K (see Fig.~\ref{Heat_capacity}). More detailed analysis is necessary to clarify the behavior of the heat capacity at higher temperatures.

%******************Figure6*****************************

The final values of the CF parameters which were used in our simulations of the $dc$ susceptibility and the magnetization for different directions of the external magnetic fields are presented in Table~\ref{CFparameters}, column B. These parameters agree qualitatively with those found in Ref.~\onlinecite{Princep2013} from the analysis of the CF excitations in the Pr$^{3+}$ ions in Pr$_{2}$Sn$_{2}$O$_{7}$ but differ markedly from the results of the studies of the crystal fields in Nd$_2$Ir$_2$O$_7$~\cite{Watahiki2011} (as far as we know, this is the only earlier published set of CF parameters for Nd$^{3+}$ ions in Nd-based pyrochlores) and Pr$_2$Zr$_2$O$_7$~\cite{Kimura2013_1} (see Table~\ref{CFparameters}). The CF interaction has been treated in Refs.~\onlinecite{Kimura2013_1} and \onlinecite{Watahiki2011} as a perturbation within the ground $J$-multiplet only. However, the CF splittings of the ground multiplets $^{4}I_{9/2}$ of Nd$^{3+}$ and $^3H_4$ of Pr$^{3+}$ ions in the pyrochlores are rather large and comparable to the excitation energies of the nearest multiplets $^4I_{11/2}$ and $^3H_5$, respectively. As a result, the $J$-mixing should be taken into account.   

The calculated energies of the sub-levels of the $^{4}I_{9/2}$ ground multiplet are 0 ($\Gamma_{56}$), 178.5 ($\Gamma_{4}$), 252 ($\Gamma_{56}$), 262 ($\Gamma_{4}$) and 785 ($\Gamma_{4}$) (in cm$^{-1}$). The corresponding irreducible representations of the $D_{3d}$ point symmetry group are presented in brackets. A transverse $g$-factor of the Kramers doublet with the $\Gamma_{56}$ symmetry  (of dipolar-octupolar type according to the classification proposed in Ref.~\onlinecite{Huang2014}) is exactly zero and our results are in line with the Ising-type magnetic anisotropy of the Nd$^{3+}$ ions in compounds with the pyrochlore structure noted previously in Ref.~\onlinecite{Yasui2003}. The calculated longitudinal $g$-factor of the ground state, $g_{\parallel} = 4.793$, is close to the value 4.6 that has been determined in Ref.~\onlinecite{Onose2004} for the Nd$^{3+}$ ions in Nd$_{2}$Mo$_{2}$O$_{7}$.

%******************Table4*****************************
\begin{table}[tb]
\caption{\label{CFparameters} Crystal field parameters $B_{p}^{q}$ (cm$^{-1}$) in some Nd and Pr based pyrochlores. The initial CF parameters for \ndzo, calculated using the exchange charge model~\cite{Klekovkina2014, Malkin2004} with the parameters of the crystal lattice determined in the present work, are given in column A. The final values of the CF parameters used to simulate the $dc$ susceptibility and the magnetization data are given in column B (see text for details).}
\centering
\begin{tabular}{ccccccc}
\hline
\hline
$p$&$q$&Pr$_{2}$Sn$_{2}$O$_{7}$&Pr$_{2}$Zr$_{2}$O$_{7}$&Nd$_{2}$Ir$_{2}$O$_{7}$&\multicolumn{2}{c}{\ndzo} \\
& &Ref.~\onlinecite{Princep2013} &Ref.~\onlinecite{Kimura2013_1} &Ref.~\onlinecite{Watahiki2011} & \multicolumn{2}{c}{Present work} \\
& & & & & A& B\\
\hline
2&0&466.3& 711& 702 & 438&460 \\
\newline
4&0&3485.5& 3173& 2110& 3096& 3200 \\
\newline
4&3&1296.6& 1093& 3087& 1027& 1132.5 \\
\newline
6&0&1163.7& 1402& 1500& 1053& 992 \\
\newline
6&3&-865.7& 235& -284& -608& -593.4 \\
\newline
6&6&1548.7& 604& 1007& 804& 1021 \\
\hline
\hline
\end{tabular}
\end{table}

The temperature dependence of the $dc$ susceptibility was calculated using the corresponding expression presented in Ref.~\onlinecite{Malkin2010}, the results of the calculations are shown in Fig.~\ref{Susceptibility}b. In order to match the experimental data, the orbital reduction factor $k$ = 0.9575 and the exchange constants $\lambda_{\parallel}$ = 0.2, $\lambda_{\perp 1}$ = -0.1 and $\lambda_{\perp 2}$ = 0 (in units of kOe/$\mu_{B}$) were introduced.

%The calculated Van Vleck susceptibility $\chi_{\mathrm{VV}}=2\sum_{k}\left(\left|\left\langle k\left|\mu_z\right|g\right\rangle\right|^2+2\left|\left\langle k\left|\mu_x\right|g\right\rangle\right|^2\right)/3\left(E_k-E_g\right)$ (here $E_k$ and $E_g$ are the energies of the excited and the ground states) of a single Nd$^{3+}$ ion in the ground state, $N_{\mathrm{A}}\chi_{\mathrm{VV}}=0.00327$~emu/mol, matches very well the value obtained from the analysis of the experimental data (see Fig.~\ref{Susceptibility} and subsection~\ref{MagHC} above). 

The calculated Van Vleck susceptibility of the Nd$^{3+}$ ions in the ground state, averaged over four magnetically non-equivalent sites in the pyrochlore lattice, 
$\chi_{\mathrm{VV}}=2\sum_{j,\alpha=x,y,z}\left|\left\langle j\left|\mu_{\alpha}\right|g\right\rangle\right|^2/3\left(E_j-E_g\right)$ (here $E_j $ and $E_g$ are the energies of the excited states $\left(j\right)$ and the ground state $\left(g\right)$), $\chi_{\mathrm{VV}}=0.00327$~~emu/(mol Nd), matches very well with the value obtained from the analysis of the experimental data (see Fig.~\ref{Susceptibility} and subsection~\ref{MagHC} above). 

Because a gap of 250~K between the ground and the first excited doublets in the energy spectrum of Nd$^{3+}$ ions is rather large, the low temperature magnetic properties of the Nd~subsystem in \ndzo\ can be considered using the truncated Hilbert space spanned by the wave functions of the ground doublets only. The projection of the exchange and magnetic dipolar interactions between the nearest neighbor Nd$^{3+}$  ions on this space can be written in terms of the pseudospin $S=1/2$ operators defined in local Cartesian coordinates with the $z$-axes along the corresponding trigonal symmetry axes crossing at the angle of $\arccos(-1/3)$: 
\begin{eqnarray} \label{eq5}
H_s=\left(J_{\mathrm{exch}}+J_{\mathrm{dd}}\right)S_{z1}S_{z2},
\end{eqnarray}                                                        
here $J_{\mathrm{exch}}=g_{||}^2\mu_{\mathrm{B}}^2\left(2\lambda_{||}-\lambda_{\bot 1}\right)/3$ and  $J_{\mathrm{dd}}=5g_{||}^2\mu_{\mathrm{B}}^2/3\left|\bm{r}_1-\bm{r}_2\right|^3$. Using the values of parameters given above, we obtain $J_{\mathrm{exch}}=0.257$~K and $J_{\mathrm{dd}}=0.442$~K. So, both the exchange and dipolar interactions between the nearest neighbor ions have antiferromagnetic character, and the exchange contribution is about half the dipolar contribution. At the first glance this result contradicts with the sign of the Weiss temperature indicating a ferromagnetic coupling. However, in case of Ising-type magnetic anisotropy along the crystal ternary axes in the pyrochlore structure, the sum of  antiferromagnetic interactions between a fixed rare-earth ion and the six nearest neighbor ions leads to formation of a local magnetic field parallel to the applied field and, correspondingly, to a positive Weiss temperature~\cite{Malkin2010}. A crude estimate, taking into account only the ground doublet of the Nd$^{3+}$ ions and neglecting the Van Vleck contribution to the susceptibility, gives a value $\theta_{\mathrm{W}}=\left(g_{||}^2\mu_{\mathrm{B}}^2/6k_{\mathrm{B}}\right)\left[2\lambda_{||}-\lambda_{\bot 1}+7.52\left(4\pi/3\nu\right)\right]=0.38$~K with the dominant contribution (0.25~K) from the dipolar interactions represented by the last term in the square brackets. According to a general form of the exchange interaction~\cite{Levy1964}, it may contain terms (neglected in the present work) constructed from single-ion operators $\sum_{f}C_2^2(f)s_+(f)$, $\sum_{f}C_{-2}^2(f)s_-(f)$, $\sum_{f}C_5^6(f)s_+(f)$, $\sum_{f}C_{-5}^6(f)s_-(f)$ (the summation over $f$ is over individual $4f$ electrons with spin $\bm{s}\left(f\right)$ on the Nd$^{3+}$ ion) which mix eigenfunctions of the pseudospin $S_z$ operator~\cite{Huang2014} and bring about fluctuations of the Ising-type magnetic moments. The effects of these terms on the low temperature magnetic properties of the Nd$^{3+}$ ions remain to be explored. It should also be noted that at temperatures below 1.5~K the single-site self-consistent approximation appears not to be valid (in particular, in contrast to the experimental data which demonstrate $\chi\left(T\right)$ doubles in the temperature range from 1 to 0.5~K, the calculated value of $\chi\left(T\right)$ increases more rapidly as the temperature decreases below 1~K).

%It should be noted that the possible uncertainties of the exchange constant values are rather large because at temperatures below 1.5~K the single-site self-consistent approximation is not valid (in particular, in contrast to to the experimental data which demonstrate $\chi(T)$ doubles in the temperature range from 0.5 to 1~K, the calculated value of $\chi(T)$ increases more rapidly as the temperature decreases from 1 to 0.5~K.

The simulated magnetic field dependence of the magnetization of \ndzo\ with the magnetic field directed along the [100], [110], and [111] symmetry axes of the pyrochlore lattice at different temperatures are compared with the experimental data in Fig.~\ref{Magn_temp}. As can be seen from Fig.~\ref{Magn_temp}, the calculated field dependences of the magnetization agree well with the experimental data.

%We believe that the obtained sets of crystal field parameters and exchange coupling constants may serve as a basis for the analysis of future experiments studying the field and temperature dependence of the magnetization at very low temperature and the interpretation of the magnetic excitation spectra. The resulting set of parameters can also be used to predict the magnetic properties of other Nd-based pyrochlore magnets.

%*************do not know if there is any point for the following paragraph****************
%The magnetic properties studies of the crystal of \ndzo reveal a ferromagnetic coupling between the Nd spins, although no magnetic transition has been observed down to 0.5 K. A spin ice-like anisotropic behavior has been observed when applying a magnetic field along different high symmetry directions.

\section{Summary}
We have investigated the structural properties of single-crystals of the pyrochlore oxide, \ndzo. Powder x-ray diffraction studies of the crystal boule confirm that the $Fd\bar{3}m$ pyrochlore phase is formed, and single-crystal x-ray diffraction and neutron diffraction data show a \ndzo\ structure with no cationic or anionic deficiencies. Supplementary investigations, such as detailed inelastic neutron scattering experiments, are now being carried out on these \ndzo\ crystals.

The magnetic susceptibility data collected reveal a spin ice-type magnetic anisotropy and an effective ferromagnetic coupling between the Nd spins. The heat capacity decreases monotonically with decreasing temperature, followed by an upturn at low temperature. However, no sign of long-range magnetic ordering was observed in either the heat capacity or the magnetization data down to 0.5~K. 

The measured temperature dependence of the magnetic susceptibility and the field dependences of the isothermal magnetization are reproduced satisfactorily by the CF calculations and subsequent modeling in the framework of the mean-field approximation taking into account the dipolar and the anisotropic exchange interactions between the Nd$^{3+}$ ions. The sets of CF parameters and exchange coupling constants obtained may serve as a basis for future studies of the spectral and magnetic properties of Nd-based pyrochlore magnets at low temperatures. 

%Supplementary investigations, such as detailed inelastic neutron scattering experiments, are now being carried out on these \ndzo\ crystals.
%In order to analyze the magnetic properties we have calculated the crystal field parameters and exchange coupling constants. The parameters of the model have been selected such as to find the crystal field parameters that best fit the experimental data for the temperature dependence of the magnetic susceptibility and the field dependence of the magnetization.\\
  
\begin{acknowledgments}
This work was supported by a grant from the EPSRC, UK (EP/I007210/1). V. V. Klekovkina and B. Z. Malkin acknowledge the support from Russian Foundation for Basic Researches (Grant N14-02-00826). The authors thank S. York for the EDAX analysis, T. E. Orton for valuable technical support, and James Hill and Alex Lee for their contribution to some of the measurements of the magnetic properties.
\end{acknowledgments}
\bibliography{Nd2Zr2O7}

\begin{thebibliography}{64}%
\makeatletter
\providecommand \@ifxundefined [1]{%
 \@ifx{#1\undefined}
}%
\providecommand \@ifnum [1]{%
 \ifnum #1\expandafter \@firstoftwo
 \else \expandafter \@secondoftwo
 \fi
}%
\providecommand \@ifx [1]{%
 \ifx #1\expandafter \@firstoftwo
 \else \expandafter \@secondoftwo
 \fi
}%
\providecommand \natexlab [1]{#1}%
\providecommand \enquote  [1]{``#1''}%
\providecommand \bibnamefont  [1]{#1}%
\providecommand \bibfnamefont [1]{#1}%
\providecommand \citenamefont [1]{#1}%
\providecommand \href@noop [0]{\@secondoftwo}%
\providecommand \href [0]{\begingroup \@sanitize@url \@href}%
\providecommand \@href[1]{\@@startlink{#1}\@@href}%
\providecommand \@@href[1]{\endgroup#1\@@endlink}%
\providecommand \@sanitize@url [0]{\catcode `\\12\catcode `\$12\catcode
  `\&12\catcode `\#12\catcode `\^12\catcode `\_12\catcode `\%12\relax}%
\providecommand \@@startlink[1]{}%
\providecommand \@@endlink[0]{}%
\providecommand \url  [0]{\begingroup\@sanitize@url \@url }%
\providecommand \@url [1]{\endgroup\@href {#1}{\urlprefix }}%
\providecommand \urlprefix  [0]{URL }%
\providecommand \Eprint [0]{\href }%
\providecommand \doibase [0]{http://dx.doi.org/}%
\providecommand \selectlanguage [0]{\@gobble}%
\providecommand \bibinfo  [0]{\@secondoftwo}%
\providecommand \bibfield  [0]{\@secondoftwo}%
\providecommand \translation [1]{[#1]}%
\providecommand \BibitemOpen [0]{}%
\providecommand \bibitemStop [0]{}%
\providecommand \bibitemNoStop [0]{.\EOS\space}%
\providecommand \EOS [0]{\spacefactor3000\relax}%
\providecommand \BibitemShut  [1]{\csname bibitem#1\endcsname}%
\let\auto@bib@innerbib\@empty
%</preamble>
\bibitem [{\citenamefont {Greedan}(2001)}]{Greedan2001}%
  \BibitemOpen
  \bibfield  {author} {\bibinfo {author} {\bibfnamefont {J.~E.}\ \bibnamefont
  {Greedan}},\ }\href@noop {} {\bibfield  {journal} {\bibinfo  {journal} {J.
  Mater. Chem.}\ }\textbf {\bibinfo {volume} {11}},\ \bibinfo {pages} {37}
  (\bibinfo {year} {2001})}\BibitemShut {NoStop}%
\bibitem [{\citenamefont {Gardner}\ \emph {et~al.}(2010)\citenamefont
  {Gardner}, \citenamefont {Gingras},\ and\ \citenamefont
  {Greedan}}]{Gardner2010}%
  \BibitemOpen
  \bibfield  {author} {\bibinfo {author} {\bibfnamefont {J.~S.}\ \bibnamefont
  {Gardner}}, \bibinfo {author} {\bibfnamefont {M.~J.~P.}\ \bibnamefont
  {Gingras}}, \ and\ \bibinfo {author} {\bibfnamefont {J.~E.}\ \bibnamefont
  {Greedan}},\ }\href@noop {} {\bibfield  {journal} {\bibinfo  {journal} {Rev.
  Mod. Phys.}\ }\textbf {\bibinfo {volume} {82}},\ \bibinfo {pages} {53}
  (\bibinfo {year} {2010})}\BibitemShut {NoStop}%
\bibitem [{\citenamefont {Harris}\ \emph {et~al.}(1997)\citenamefont {Harris},
  \citenamefont {Bramwell}, \citenamefont {McMorrow}, \citenamefont {Zeiske},\
  and\ \citenamefont {Godfrey}}]{Harris1997}%
  \BibitemOpen
  \bibfield  {author} {\bibinfo {author} {\bibfnamefont {M.~J.}\ \bibnamefont
  {Harris}}, \bibinfo {author} {\bibfnamefont {S.~T.}\ \bibnamefont
  {Bramwell}}, \bibinfo {author} {\bibfnamefont {D.~F.}\ \bibnamefont
  {McMorrow}}, \bibinfo {author} {\bibfnamefont {T.}~\bibnamefont {Zeiske}}, \
  and\ \bibinfo {author} {\bibfnamefont {K.~W.}\ \bibnamefont {Godfrey}},\
  }\href@noop {} {\bibfield  {journal} {\bibinfo  {journal} {Phys. Rev. Lett.}\
  }\textbf {\bibinfo {volume} {79}},\ \bibinfo {pages} {2554} (\bibinfo {year}
  {1997})}\BibitemShut {NoStop}%
\bibitem [{\citenamefont {Ramirez}\ \emph {et~al.}(1999)\citenamefont
  {Ramirez}, \citenamefont {Hayashi}, \citenamefont {Cava}, \citenamefont
  {Siddharthan},\ and\ \citenamefont {Shastry}}]{Ramirez1999}%
  \BibitemOpen
  \bibfield  {author} {\bibinfo {author} {\bibfnamefont {A.~P.}\ \bibnamefont
  {Ramirez}}, \bibinfo {author} {\bibfnamefont {A.}~\bibnamefont {Hayashi}},
  \bibinfo {author} {\bibfnamefont {R.~J.}\ \bibnamefont {Cava}}, \bibinfo
  {author} {\bibfnamefont {R.}~\bibnamefont {Siddharthan}}, \ and\ \bibinfo
  {author} {\bibfnamefont {B.~S.}\ \bibnamefont {Shastry}},\ }\href@noop {}
  {\bibfield  {journal} {\bibinfo  {journal} {Nature}\ }\textbf {\bibinfo
  {volume} {399}},\ \bibinfo {pages} {333} (\bibinfo {year}
  {1999})}\BibitemShut {NoStop}%
\bibitem [{\citenamefont {Gaulin}\ \emph {et~al.}(1992)\citenamefont {Gaulin},
  \citenamefont {Reimers}, \citenamefont {Mason}, \citenamefont {Greedan},\
  and\ \citenamefont {Tun}}]{Gaulin1992}%
  \BibitemOpen
  \bibfield  {author} {\bibinfo {author} {\bibfnamefont {B.~D.}\ \bibnamefont
  {Gaulin}}, \bibinfo {author} {\bibfnamefont {J.~N.}\ \bibnamefont {Reimers}},
  \bibinfo {author} {\bibfnamefont {T.~E.}\ \bibnamefont {Mason}}, \bibinfo
  {author} {\bibfnamefont {J.~E.}\ \bibnamefont {Greedan}}, \ and\ \bibinfo
  {author} {\bibfnamefont {Z.}~\bibnamefont {Tun}},\ }\href@noop {} {\bibfield
  {journal} {\bibinfo  {journal} {Phys. Rev. Lett.}\ }\textbf {\bibinfo
  {volume} {69}},\ \bibinfo {pages} {3244} (\bibinfo {year}
  {1992})}\BibitemShut {NoStop}%
\bibitem [{\citenamefont {Gardner}\ \emph
  {et~al.}(1999{\natexlab{a}})\citenamefont {Gardner}, \citenamefont {Gaulin},
  \citenamefont {Lee}, \citenamefont {Broholm}, \citenamefont {Raju},\ and\
  \citenamefont {Greedan}}]{Gardner1999_1}%
  \BibitemOpen
  \bibfield  {author} {\bibinfo {author} {\bibfnamefont {J.~S.}\ \bibnamefont
  {Gardner}}, \bibinfo {author} {\bibfnamefont {B.~D.}\ \bibnamefont {Gaulin}},
  \bibinfo {author} {\bibfnamefont {S.-H.}\ \bibnamefont {Lee}}, \bibinfo
  {author} {\bibfnamefont {C.}~\bibnamefont {Broholm}}, \bibinfo {author}
  {\bibfnamefont {N.~P.}\ \bibnamefont {Raju}}, \ and\ \bibinfo {author}
  {\bibfnamefont {J.~E.}\ \bibnamefont {Greedan}},\ }\href@noop {} {\bibfield
  {journal} {\bibinfo  {journal} {Phys. Rev. Lett.}\ }\textbf {\bibinfo
  {volume} {83}},\ \bibinfo {pages} {211} (\bibinfo {year}
  {1999}{\natexlab{a}})}\BibitemShut {NoStop}%
\bibitem [{\citenamefont {Gardner}\ \emph
  {et~al.}(1999{\natexlab{b}})\citenamefont {Gardner}, \citenamefont
  {Dunsiger}, \citenamefont {Gaulin}, \citenamefont {Gingras}, \citenamefont
  {Greedan}, \citenamefont {Kiefl}, \citenamefont {Lumsden}, \citenamefont
  {MacFarlane}, \citenamefont {Raju}, \citenamefont {Sonier}, \citenamefont
  {Swainson},\ and\ \citenamefont {Tun}}]{Gardner1999_2}%
  \BibitemOpen
  \bibfield  {author} {\bibinfo {author} {\bibfnamefont {J.~S.}\ \bibnamefont
  {Gardner}}, \bibinfo {author} {\bibfnamefont {S.~R.}\ \bibnamefont
  {Dunsiger}}, \bibinfo {author} {\bibfnamefont {B.~D.}\ \bibnamefont
  {Gaulin}}, \bibinfo {author} {\bibfnamefont {M.~J.~P.}\ \bibnamefont
  {Gingras}}, \bibinfo {author} {\bibfnamefont {J.~E.}\ \bibnamefont
  {Greedan}}, \bibinfo {author} {\bibfnamefont {R.~F.}\ \bibnamefont {Kiefl}},
  \bibinfo {author} {\bibfnamefont {M.~D.}\ \bibnamefont {Lumsden}}, \bibinfo
  {author} {\bibfnamefont {W.~A.}\ \bibnamefont {MacFarlane}}, \bibinfo
  {author} {\bibfnamefont {N.~P.}\ \bibnamefont {Raju}}, \bibinfo {author}
  {\bibfnamefont {J.~E.}\ \bibnamefont {Sonier}}, \bibinfo {author}
  {\bibfnamefont {I.}~\bibnamefont {Swainson}}, \ and\ \bibinfo {author}
  {\bibfnamefont {Z.}~\bibnamefont {Tun}},\ }\href@noop {} {\bibfield
  {journal} {\bibinfo  {journal} {Phys. Rev. Lett.}\ }\textbf {\bibinfo
  {volume} {82}},\ \bibinfo {pages} {1012} (\bibinfo {year}
  {1999}{\natexlab{b}})}\BibitemShut {NoStop}%
\bibitem [{\citenamefont {Molavian}\ \emph {et~al.}(2007)\citenamefont
  {Molavian}, \citenamefont {Gingras},\ and\ \citenamefont
  {Canals}}]{Molavian2007}%
  \BibitemOpen
  \bibfield  {author} {\bibinfo {author} {\bibfnamefont {H.~R.}\ \bibnamefont
  {Molavian}}, \bibinfo {author} {\bibfnamefont {M.~J.~P.}\ \bibnamefont
  {Gingras}}, \ and\ \bibinfo {author} {\bibfnamefont {B.}~\bibnamefont
  {Canals}},\ }\href@noop {} {\bibfield  {journal} {\bibinfo  {journal} {Phys.
  Rev. Lett.}\ }\textbf {\bibinfo {volume} {98}},\ \bibinfo {pages} {157204}
  (\bibinfo {year} {2007})}\BibitemShut {NoStop}%
\bibitem [{\citenamefont {Ross}\ \emph {et~al.}(2011)\citenamefont {Ross},
  \citenamefont {Savary}, \citenamefont {Gaulin},\ and\ \citenamefont
  {Balents}}]{Ross2011}%
  \BibitemOpen
  \bibfield  {author} {\bibinfo {author} {\bibfnamefont {K.~A.}\ \bibnamefont
  {Ross}}, \bibinfo {author} {\bibfnamefont {L.}~\bibnamefont {Savary}},
  \bibinfo {author} {\bibfnamefont {B.~D.}\ \bibnamefont {Gaulin}}, \ and\
  \bibinfo {author} {\bibfnamefont {L.}~\bibnamefont {Balents}},\ }\href@noop
  {} {\bibfield  {journal} {\bibinfo  {journal} {Phys. Rev. X}\ }\textbf
  {\bibinfo {volume} {1}},\ \bibinfo {pages} {021002} (\bibinfo {year}
  {2011})}\BibitemShut {NoStop}%
\bibitem [{\citenamefont {Gingras}\ and\ \citenamefont
  {McClarty}(2014)}]{Gingras2014}%
  \BibitemOpen
  \bibfield  {author} {\bibinfo {author} {\bibfnamefont {M.~J.~P.}\
  \bibnamefont {Gingras}}\ and\ \bibinfo {author} {\bibfnamefont {P.~A.}\
  \bibnamefont {McClarty}},\ }\href@noop {} {\bibfield  {journal} {\bibinfo
  {journal} {Rep. Prog. Phys.}\ }\textbf {\bibinfo {volume} {77}},\ \bibinfo
  {pages} {056501} (\bibinfo {year} {2014})}\BibitemShut {NoStop}%
\bibitem [{\citenamefont {Balakrishnan}\ \emph {et~al.}(1998)\citenamefont
  {Balakrishnan}, \citenamefont {Petrenko}, \citenamefont {Lees},\ and\
  \citenamefont {Paul}}]{Balakrishnan1998}%
  \BibitemOpen
  \bibfield  {author} {\bibinfo {author} {\bibfnamefont {G.}~\bibnamefont
  {Balakrishnan}}, \bibinfo {author} {\bibfnamefont {O.~A.}\ \bibnamefont
  {Petrenko}}, \bibinfo {author} {\bibfnamefont {M.~R.}\ \bibnamefont {Lees}},
  \ and\ \bibinfo {author} {\bibfnamefont {D.~M.}\ \bibnamefont {Paul}},\
  }\href@noop {} {\bibfield  {journal} {\bibinfo  {journal} {J. Phys.: Condens.
  Matter.}\ }\textbf {\bibinfo {volume} {10}},\ \bibinfo {pages} {L723}
  (\bibinfo {year} {1998})}\BibitemShut {NoStop}%
\bibitem [{\citenamefont {Gardner}\ \emph {et~al.}(1998)\citenamefont
  {Gardner}, \citenamefont {Gaulin},\ and\ \citenamefont {Paul}}]{Gardner98}%
  \BibitemOpen
  \bibfield  {author} {\bibinfo {author} {\bibfnamefont {J.~S.}\ \bibnamefont
  {Gardner}}, \bibinfo {author} {\bibfnamefont {B.~D.}\ \bibnamefont {Gaulin}},
  \ and\ \bibinfo {author} {\bibfnamefont {D.~M.}\ \bibnamefont {Paul}},\
  }\href@noop {} {\bibfield  {journal} {\bibinfo  {journal} {{J. Cryst.
  Growth}}\ }\textbf {\bibinfo {volume} {{191}}},\ \bibinfo {pages} {{740}}
  (\bibinfo {year} {{1998}})}\BibitemShut {NoStop}%
\bibitem [{\citenamefont {Prabhakaran}\ and\ \citenamefont
  {Boothroyd}(2011)}]{Prabhakaran11}%
  \BibitemOpen
  \bibfield  {author} {\bibinfo {author} {\bibfnamefont {D.}~\bibnamefont
  {Prabhakaran}}\ and\ \bibinfo {author} {\bibfnamefont {A.~T.}\ \bibnamefont
  {Boothroyd}},\ }\href@noop {} {\bibfield  {journal} {\bibinfo  {journal} {{J.
  Cryst. Growth}}\ }\textbf {\bibinfo {volume} {{318}}},\ \bibinfo {pages}
  {{1053}} (\bibinfo {year} {{2011}})}\BibitemShut {NoStop}%
\bibitem [{oth()}]{otheruses}%
  \BibitemOpen
  \href@noop {} {}\bibinfo {note} {The research community has also recently
  shown an increased interest in the lanthanide zirconates,
  $A_{2}$Zr$_{2}$O$_{7}$, due to their potential use in the immobilization of
  radioactive waste~\cite{Ewing2004} and in thermal barrier
  coatings~\cite{Wu2002}.}\BibitemShut {Stop}%
\bibitem [{\citenamefont {Michel}\ \emph {et~al.}(1974)\citenamefont {Michel},
  \citenamefont {{Perez y Jorba}},\ and\ \citenamefont
  {Collongues}}]{Michel1974}%
  \BibitemOpen
  \bibfield  {author} {\bibinfo {author} {\bibfnamefont {D.}~\bibnamefont
  {Michel}}, \bibinfo {author} {\bibfnamefont {M.}~\bibnamefont {{Perez y
  Jorba}}}, \ and\ \bibinfo {author} {\bibfnamefont {R.}~\bibnamefont
  {Collongues}},\ }\href@noop {} {\bibfield  {journal} {\bibinfo  {journal}
  {Mat. Res. Bull.}\ }\textbf {\bibinfo {volume} {9}},\ \bibinfo {pages} {1457}
  (\bibinfo {year} {1974})}\BibitemShut {NoStop}%
\bibitem [{\citenamefont {Blanchard}\ \emph {et~al.}(2012)\citenamefont
  {Blanchard}, \citenamefont {Clements}, \citenamefont {Kennedy}, \citenamefont
  {Ling}, \citenamefont {Reynolds}, \citenamefont {Avdeev}, \citenamefont
  {Stampfl}, \citenamefont {Zhang},\ and\ \citenamefont
  {Jang}}]{Blanchard2012}%
  \BibitemOpen
  \bibfield  {author} {\bibinfo {author} {\bibfnamefont {P.~E.~R.}\
  \bibnamefont {Blanchard}}, \bibinfo {author} {\bibfnamefont {R.}~\bibnamefont
  {Clements}}, \bibinfo {author} {\bibfnamefont {B.~J.}\ \bibnamefont
  {Kennedy}}, \bibinfo {author} {\bibfnamefont {C.~D.}\ \bibnamefont {Ling}},
  \bibinfo {author} {\bibfnamefont {E.}~\bibnamefont {Reynolds}}, \bibinfo
  {author} {\bibfnamefont {M.}~\bibnamefont {Avdeev}}, \bibinfo {author}
  {\bibfnamefont {A.~P.~J.}\ \bibnamefont {Stampfl}}, \bibinfo {author}
  {\bibfnamefont {Z.}~\bibnamefont {Zhang}}, \ and\ \bibinfo {author}
  {\bibfnamefont {L.-Y.}\ \bibnamefont {Jang}},\ }\href@noop {} {\bibfield
  {journal} {\bibinfo  {journal} {Inorg. Chem.}\ }\textbf {\bibinfo {volume}
  {51}},\ \bibinfo {pages} {13237} (\bibinfo {year} {2012})}\BibitemShut
  {NoStop}%
\bibitem [{\citenamefont {Rushton}\ \emph {et~al.}(2004)\citenamefont
  {Rushton}, \citenamefont {Grimes}, \citenamefont {Stanek},\ and\
  \citenamefont {Owens}}]{Rushton2004}%
  \BibitemOpen
  \bibfield  {author} {\bibinfo {author} {\bibfnamefont {M.~J.~D.}\
  \bibnamefont {Rushton}}, \bibinfo {author} {\bibfnamefont {R.~W.}\
  \bibnamefont {Grimes}}, \bibinfo {author} {\bibfnamefont {C.~R.}\
  \bibnamefont {Stanek}}, \ and\ \bibinfo {author} {\bibfnamefont
  {S.}~\bibnamefont {Owens}},\ }\href@noop {} {\bibfield  {journal} {\bibinfo
  {journal} {J. Mater. Res.}\ }\textbf {\bibinfo {volume} {19}},\ \bibinfo
  {pages} {1603} (\bibinfo {year} {2004})}\BibitemShut {NoStop}%
\bibitem [{\citenamefont {Clements}\ \emph {et~al.}(2011)\citenamefont
  {Clements}, \citenamefont {Hester}, \citenamefont {Kennedy}, \citenamefont
  {Ling},\ and\ \citenamefont {Stampfl}}]{Clements2011}%
  \BibitemOpen
  \bibfield  {author} {\bibinfo {author} {\bibfnamefont {R.}~\bibnamefont
  {Clements}}, \bibinfo {author} {\bibfnamefont {J.~R.}\ \bibnamefont
  {Hester}}, \bibinfo {author} {\bibfnamefont {B.~J.}\ \bibnamefont {Kennedy}},
  \bibinfo {author} {\bibfnamefont {C.~D.}\ \bibnamefont {Ling}}, \ and\
  \bibinfo {author} {\bibfnamefont {A.~J.~P.}\ \bibnamefont {Stampfl}},\
  }\href@noop {} {\bibfield  {journal} {\bibinfo  {journal} {J. Solid State
  Chem.}\ }\textbf {\bibinfo {volume} {184}},\ \bibinfo {pages} {2108}
  (\bibinfo {year} {2011})}\BibitemShut {NoStop}%
\bibitem [{\citenamefont {Blanchard}\ \emph {et~al.}(2013)\citenamefont
  {Blanchard}, \citenamefont {Liu}, \citenamefont {Kennedy}, \citenamefont
  {Ling}, \citenamefont {Zhang}, \citenamefont {Avdeev}, \citenamefont {Cowie},
  \citenamefont {Thomsen},\ and\ \citenamefont {Jang}}]{Blanchard2013}%
  \BibitemOpen
  \bibfield  {author} {\bibinfo {author} {\bibfnamefont {P.~E.~R.}\
  \bibnamefont {Blanchard}}, \bibinfo {author} {\bibfnamefont {S.}~\bibnamefont
  {Liu}}, \bibinfo {author} {\bibfnamefont {B.~J.}\ \bibnamefont {Kennedy}},
  \bibinfo {author} {\bibfnamefont {C.~D.}\ \bibnamefont {Ling}}, \bibinfo
  {author} {\bibfnamefont {Z.}~\bibnamefont {Zhang}}, \bibinfo {author}
  {\bibfnamefont {M.}~\bibnamefont {Avdeev}}, \bibinfo {author} {\bibfnamefont
  {B.~C.~C.}\ \bibnamefont {Cowie}}, \bibinfo {author} {\bibfnamefont
  {L.}~\bibnamefont {Thomsen}}, \ and\ \bibinfo {author} {\bibfnamefont
  {L.-Y.}\ \bibnamefont {Jang}},\ }\href@noop {} {\bibfield  {journal}
  {\bibinfo  {journal} {Dalton Trans.}\ }\textbf {\bibinfo {volume} {42}},\
  \bibinfo {pages} {14875} (\bibinfo {year} {2013})}\BibitemShut {NoStop}%
\bibitem [{\citenamefont {Bl\"{o}te}\ \emph {et~al.}(1969)\citenamefont
  {Bl\"{o}te}, \citenamefont {Wielinga},\ and\ \citenamefont
  {Huiskamp}}]{Blote1969}%
  \BibitemOpen
  \bibfield  {author} {\bibinfo {author} {\bibfnamefont {H.~W.~J.}\
  \bibnamefont {Bl\"{o}te}}, \bibinfo {author} {\bibfnamefont {R.~F.}\
  \bibnamefont {Wielinga}}, \ and\ \bibinfo {author} {\bibfnamefont {W.~J.}\
  \bibnamefont {Huiskamp}},\ }\href@noop {} {\bibfield  {journal} {\bibinfo
  {journal} {Physica}\ }\textbf {\bibinfo {volume} {43}},\ \bibinfo {pages}
  {549} (\bibinfo {year} {1969})}\BibitemShut {NoStop}%
\bibitem [{\citenamefont {Matsuhira}\ \emph {et~al.}(2009)\citenamefont
  {Matsuhira}, \citenamefont {Sekine}, \citenamefont {Paulsen}, \citenamefont
  {Wakeshima}, \citenamefont {Hinatsu}, \citenamefont {Kitazawa}, \citenamefont
  {Kiuchi}, \citenamefont {Hiroi},\ and\ \citenamefont
  {Takagi}}]{Matsuhira2009}%
  \BibitemOpen
  \bibfield  {author} {\bibinfo {author} {\bibfnamefont {K.}~\bibnamefont
  {Matsuhira}}, \bibinfo {author} {\bibfnamefont {C.}~\bibnamefont {Sekine}},
  \bibinfo {author} {\bibfnamefont {C.}~\bibnamefont {Paulsen}}, \bibinfo
  {author} {\bibfnamefont {M.}~\bibnamefont {Wakeshima}}, \bibinfo {author}
  {\bibfnamefont {Y.}~\bibnamefont {Hinatsu}}, \bibinfo {author} {\bibfnamefont
  {T.}~\bibnamefont {Kitazawa}}, \bibinfo {author} {\bibfnamefont
  {Y.}~\bibnamefont {Kiuchi}}, \bibinfo {author} {\bibfnamefont
  {Z.}~\bibnamefont {Hiroi}}, \ and\ \bibinfo {author} {\bibfnamefont
  {S.}~\bibnamefont {Takagi}},\ }\href@noop {} {\bibfield  {journal} {\bibinfo
  {journal} {J. Phys.: Conf. Ser.}\ }\textbf {\bibinfo {volume} {145}},\
  \bibinfo {pages} {012031} (\bibinfo {year} {2009})}\BibitemShut {NoStop}%
\bibitem [{\citenamefont {{Ciomaga Hatnean}}\ \emph
  {et~al.}(2014{\natexlab{a}})\citenamefont {{Ciomaga Hatnean}}, \citenamefont
  {Decorse}, \citenamefont {Lees}, \citenamefont {Petrenko}, \citenamefont
  {Keeble},\ and\ \citenamefont {Balakrishnan}}]{Hatnean2014_1}%
  \BibitemOpen
  \bibfield  {author} {\bibinfo {author} {\bibfnamefont {M.}~\bibnamefont
  {{Ciomaga Hatnean}}}, \bibinfo {author} {\bibfnamefont {C.}~\bibnamefont
  {Decorse}}, \bibinfo {author} {\bibfnamefont {M.~R.}\ \bibnamefont {Lees}},
  \bibinfo {author} {\bibfnamefont {O.~A.}\ \bibnamefont {Petrenko}}, \bibinfo
  {author} {\bibfnamefont {D.~S.}\ \bibnamefont {Keeble}}, \ and\ \bibinfo
  {author} {\bibfnamefont {G.}~\bibnamefont {Balakrishnan}},\ }\href@noop {}
  {\bibfield  {journal} {\bibinfo  {journal} {Mater. Res. Express}\ }\textbf
  {\bibinfo {volume} {1}},\ \bibinfo {pages} {026109} (\bibinfo {year}
  {2014}{\natexlab{a}})}\BibitemShut {NoStop}%
\bibitem [{\citenamefont {Koohpayeh}\ \emph {et~al.}(2014)\citenamefont
  {Koohpayeh}, \citenamefont {Wen}, \citenamefont {Trump}, \citenamefont
  {Broholm},\ and\ \citenamefont {McQueen}}]{Koohpayeh2014}%
  \BibitemOpen
  \bibfield  {author} {\bibinfo {author} {\bibfnamefont {S.~M.}\ \bibnamefont
  {Koohpayeh}}, \bibinfo {author} {\bibfnamefont {J.-J.}\ \bibnamefont {Wen}},
  \bibinfo {author} {\bibfnamefont {B.~A.}\ \bibnamefont {Trump}}, \bibinfo
  {author} {\bibfnamefont {C.~L.}\ \bibnamefont {Broholm}}, \ and\ \bibinfo
  {author} {\bibfnamefont {T.~M.}\ \bibnamefont {McQueen}},\ }\href@noop {}
  {\bibfield  {journal} {\bibinfo  {journal} {J. Cryst. Growth}\ }\textbf
  {\bibinfo {volume} {402}},\ \bibinfo {pages} {291} (\bibinfo {year}
  {2014})}\BibitemShut {NoStop}%
\bibitem [{\citenamefont {Kimura}\ \emph
  {et~al.}(2013{\natexlab{a}})\citenamefont {Kimura}, \citenamefont
  {Nakatsuji}, \citenamefont {Wen}, \citenamefont {Broholm}, \citenamefont
  {Stone}, \citenamefont {Nishibori},\ and\ \citenamefont
  {Sawa}}]{Kimura2013_1}%
  \BibitemOpen
  \bibfield  {author} {\bibinfo {author} {\bibfnamefont {K.}~\bibnamefont
  {Kimura}}, \bibinfo {author} {\bibfnamefont {S.}~\bibnamefont {Nakatsuji}},
  \bibinfo {author} {\bibfnamefont {J.-J.}\ \bibnamefont {Wen}}, \bibinfo
  {author} {\bibfnamefont {C.}~\bibnamefont {Broholm}}, \bibinfo {author}
  {\bibfnamefont {M.~B.}\ \bibnamefont {Stone}}, \bibinfo {author}
  {\bibfnamefont {E.}~\bibnamefont {Nishibori}}, \ and\ \bibinfo {author}
  {\bibfnamefont {H.}~\bibnamefont {Sawa}},\ }\href@noop {} {\bibfield
  {journal} {\bibinfo  {journal} {Nat. Commun.}\ }\textbf {\bibinfo {volume}
  {4}},\ \bibinfo {pages} {1934} (\bibinfo {year}
  {2013}{\natexlab{a}})}\BibitemShut {NoStop}%
\bibitem [{\citenamefont {Kimura}\ \emph
  {et~al.}(2013{\natexlab{b}})\citenamefont {Kimura}, \citenamefont
  {Nakatsuji},\ and\ \citenamefont {Nugroho}}]{Kimura2013_2}%
  \BibitemOpen
  \bibfield  {author} {\bibinfo {author} {\bibfnamefont {K.}~\bibnamefont
  {Kimura}}, \bibinfo {author} {\bibfnamefont {S.}~\bibnamefont {Nakatsuji}}, \
  and\ \bibinfo {author} {\bibfnamefont {A.~A.}\ \bibnamefont {Nugroho}},\
  }\href@noop {} {\bibfield  {journal} {\bibinfo  {journal} {J. Korean Phys.
  Soc.}\ }\textbf {\bibinfo {volume} {63}},\ \bibinfo {pages} {719} (\bibinfo
  {year} {2013}{\natexlab{b}})}\BibitemShut {NoStop}%
\bibitem [{\citenamefont {Onoda}\ and\ \citenamefont
  {Tanaka}(2011)}]{Onoda2011}%
  \BibitemOpen
  \bibfield  {author} {\bibinfo {author} {\bibfnamefont {S.}~\bibnamefont
  {Onoda}}\ and\ \bibinfo {author} {\bibfnamefont {Y.}~\bibnamefont {Tanaka}},\
  }\href@noop {} {\bibfield  {journal} {\bibinfo  {journal} {Phys. Rev. B}\
  }\textbf {\bibinfo {volume} {83}},\ \bibinfo {pages} {094411} (\bibinfo
  {year} {2011})}\BibitemShut {NoStop}%
\bibitem [{\citenamefont {Lee}\ \emph {et~al.}(2012)\citenamefont {Lee},
  \citenamefont {Onoda},\ and\ \citenamefont {Balents}}]{Lee2012}%
  \BibitemOpen
  \bibfield  {author} {\bibinfo {author} {\bibfnamefont {S.~B.}\ \bibnamefont
  {Lee}}, \bibinfo {author} {\bibfnamefont {S.}~\bibnamefont {Onoda}}, \ and\
  \bibinfo {author} {\bibfnamefont {L.}~\bibnamefont {Balents}},\ }\href@noop
  {} {\bibfield  {journal} {\bibinfo  {journal} {Phys. Rev. B}\ }\textbf
  {\bibinfo {volume} {86}},\ \bibinfo {pages} {104412} (\bibinfo {year}
  {2012})}\BibitemShut {NoStop}%
\bibitem [{\citenamefont {{van Duijn}}\ \emph {et~al.}()\citenamefont {{van
  Duijn}}, \citenamefont {Kim}, \citenamefont {Hur}, \citenamefont {Adroja},
  \citenamefont {Bridges}, \citenamefont {Daoud-Aladine}, \citenamefont
  {Fernandez-Alonso}, \citenamefont {Ruiz-Bustos}, \citenamefont {Wen},
  \citenamefont {Kearney}, \citenamefont {Huang}, \citenamefont {Cheong},
  \citenamefont {Nakatsuji}, \citenamefont {Broholm},\ and\ \citenamefont
  {Perring}}]{Duijn14}%
  \BibitemOpen
  \bibfield  {author} {\bibinfo {author} {\bibfnamefont {J.}~\bibnamefont {{van
  Duijn}}}, \bibinfo {author} {\bibfnamefont {K.~H.}\ \bibnamefont {Kim}},
  \bibinfo {author} {\bibfnamefont {N.}~\bibnamefont {Hur}}, \bibinfo {author}
  {\bibfnamefont {D.~T.}\ \bibnamefont {Adroja}}, \bibinfo {author}
  {\bibfnamefont {F.}~\bibnamefont {Bridges}}, \bibinfo {author} {\bibfnamefont
  {A.}~\bibnamefont {Daoud-Aladine}}, \bibinfo {author} {\bibfnamefont
  {F.}~\bibnamefont {Fernandez-Alonso}}, \bibinfo {author} {\bibfnamefont
  {R.}~\bibnamefont {Ruiz-Bustos}}, \bibinfo {author} {\bibfnamefont
  {J.}~\bibnamefont {Wen}}, \bibinfo {author} {\bibfnamefont {V.}~\bibnamefont
  {Kearney}}, \bibinfo {author} {\bibfnamefont {Q.~Z.}\ \bibnamefont {Huang}},
  \bibinfo {author} {\bibfnamefont {S.-W.}\ \bibnamefont {Cheong}}, \bibinfo
  {author} {\bibfnamefont {S.}~\bibnamefont {Nakatsuji}}, \bibinfo {author}
  {\bibfnamefont {C.}~\bibnamefont {Broholm}}, \ and\ \bibinfo {author}
  {\bibfnamefont {T.~G.}\ \bibnamefont {Perring}},\ }\href@noop {} {}\bibinfo
  {note} {{a}rXiv 1407.0661}\BibitemShut {NoStop}%
\bibitem [{\citenamefont {Roth}(1956)}]{Roth1956}%
  \BibitemOpen
  \bibfield  {author} {\bibinfo {author} {\bibfnamefont {R.~S.}\ \bibnamefont
  {Roth}},\ }\href@noop {} {\bibfield  {journal} {\bibinfo  {journal} {J. Res.
  Natl. Bur. Stand.}\ }\textbf {\bibinfo {volume} {56}},\ \bibinfo {pages}
  {2643} (\bibinfo {year} {1956})}\BibitemShut {NoStop}%
\bibitem [{\citenamefont {Ohtani}\ \emph {et~al.}(2005)\citenamefont {Ohtani},
  \citenamefont {Matsumoto}, \citenamefont {Sundman}, \citenamefont {Sakuma},\
  and\ \citenamefont {Hasebe}}]{Ohtani2005}%
  \BibitemOpen
  \bibfield  {author} {\bibinfo {author} {\bibfnamefont {H.}~\bibnamefont
  {Ohtani}}, \bibinfo {author} {\bibfnamefont {S.}~\bibnamefont {Matsumoto}},
  \bibinfo {author} {\bibfnamefont {B.}~\bibnamefont {Sundman}}, \bibinfo
  {author} {\bibfnamefont {T.}~\bibnamefont {Sakuma}}, \ and\ \bibinfo {author}
  {\bibfnamefont {M.}~\bibnamefont {Hasebe}},\ }\href@noop {} {\bibfield
  {journal} {\bibinfo  {journal} {Mater. Trans.}\ }\textbf {\bibinfo {volume}
  {46}},\ \bibinfo {pages} {1167} (\bibinfo {year} {2005})}\BibitemShut
  {NoStop}%
\bibitem [{\citenamefont {Payne}\ \emph {et~al.}(2013)\citenamefont {Payne},
  \citenamefont {Tucker},\ and\ \citenamefont {Evans}}]{Payne2013}%
  \BibitemOpen
  \bibfield  {author} {\bibinfo {author} {\bibfnamefont {J.~L.}\ \bibnamefont
  {Payne}}, \bibinfo {author} {\bibfnamefont {M.~G.}\ \bibnamefont {Tucker}}, \
  and\ \bibinfo {author} {\bibfnamefont {I.~R.}\ \bibnamefont {Evans}},\
  }\href@noop {} {\bibfield  {journal} {\bibinfo  {journal} {J. Solid State
  Chem.}\ }\textbf {\bibinfo {volume} {205}},\ \bibinfo {pages} {29} (\bibinfo
  {year} {2013})}\BibitemShut {NoStop}%
\bibitem [{\citenamefont {{Ciomaga Hatnean}}\ \emph {et~al.}(2015)\citenamefont
  {{Ciomaga Hatnean}}, \citenamefont {Lees},\ and\ \citenamefont
  {Balakrishnan}}]{Hatnean2014_2}%
  \BibitemOpen
  \bibfield  {author} {\bibinfo {author} {\bibfnamefont {M.}~\bibnamefont
  {{Ciomaga Hatnean}}}, \bibinfo {author} {\bibfnamefont {M.~R.}\ \bibnamefont
  {Lees}}, \ and\ \bibinfo {author} {\bibfnamefont {G.}~\bibnamefont
  {Balakrishnan}},\ }\href@noop {} {\bibfield  {journal} {\bibinfo  {journal}
  {J. Cryst. Growth}\ }\textbf {\bibinfo {volume} {418}},\ \bibinfo {pages} {1}
  (\bibinfo {year} {2015})}\BibitemShut {NoStop}%
\bibitem [{\citenamefont {Sheldrick}(2008)}]{Sheldrick2008}%
  \BibitemOpen
  \bibfield  {author} {\bibinfo {author} {\bibfnamefont {G.~M.}\ \bibnamefont
  {Sheldrick}},\ }\href@noop {} {\bibfield  {journal} {\bibinfo  {journal}
  {Acta Cryst.}\ }\textbf {\bibinfo {volume} {A64}},\ \bibinfo {pages} {112}
  (\bibinfo {year} {2008})}\BibitemShut {NoStop}%
\bibitem [{\citenamefont {Dolomanov}\ \emph {et~al.}(2009)\citenamefont
  {Dolomanov}, \citenamefont {Bourhis}, \citenamefont {Gildea}, \citenamefont
  {Howard},\ and\ \citenamefont {Puschmann}}]{Dolomanov2009}%
  \BibitemOpen
  \bibfield  {author} {\bibinfo {author} {\bibfnamefont {O.~V.}\ \bibnamefont
  {Dolomanov}}, \bibinfo {author} {\bibfnamefont {L.~J.}\ \bibnamefont
  {Bourhis}}, \bibinfo {author} {\bibfnamefont {R.~J.}\ \bibnamefont {Gildea}},
  \bibinfo {author} {\bibfnamefont {J.~A.~K.}\ \bibnamefont {Howard}}, \ and\
  \bibinfo {author} {\bibfnamefont {H.}~\bibnamefont {Puschmann}},\ }\href@noop
  {} {\bibfield  {journal} {\bibinfo  {journal} {J. Appl. Cryst.}\ }\textbf
  {\bibinfo {volume} {42}},\ \bibinfo {pages} {339} (\bibinfo {year}
  {2009})}\BibitemShut {NoStop}%
\bibitem [{\citenamefont {Keen}\ \emph {et~al.}(2006)\citenamefont {Keen},
  \citenamefont {Gutmann},\ and\ \citenamefont {Wilson}}]{Keen2006}%
  \BibitemOpen
  \bibfield  {author} {\bibinfo {author} {\bibfnamefont {D.~A.}\ \bibnamefont
  {Keen}}, \bibinfo {author} {\bibfnamefont {M.~J.}\ \bibnamefont {Gutmann}}, \
  and\ \bibinfo {author} {\bibfnamefont {C.~C.}\ \bibnamefont {Wilson}},\
  }\href@noop {} {\bibfield  {journal} {\bibinfo  {journal} {J. Appl.
  Crystallogr.}\ }\textbf {\bibinfo {volume} {39}},\ \bibinfo {pages} {714}
  (\bibinfo {year} {2006})}\BibitemShut {NoStop}%
\bibitem [{\citenamefont {Aharoni}(1998)}]{Aharoni1998}%
  \BibitemOpen
  \bibfield  {author} {\bibinfo {author} {\bibfnamefont {A.}~\bibnamefont
  {Aharoni}},\ }\href@noop {} {\bibfield  {journal} {\bibinfo  {journal} {J.
  Appl. Phys.}\ }\textbf {\bibinfo {volume} {83}},\ \bibinfo {pages} {3432}
  (\bibinfo {year} {1998})}\BibitemShut {NoStop}%
\bibitem [{\citenamefont {Subramanian}\ \emph {et~al.}(1983)\citenamefont
  {Subramanian}, \citenamefont {Aravamudan},\ and\ \citenamefont
  {Rao}}]{Subramanian1983}%
  \BibitemOpen
  \bibfield  {author} {\bibinfo {author} {\bibfnamefont {M.~A.}\ \bibnamefont
  {Subramanian}}, \bibinfo {author} {\bibfnamefont {G.}~\bibnamefont
  {Aravamudan}}, \ and\ \bibinfo {author} {\bibfnamefont {G.~V.~S.}\
  \bibnamefont {Rao}},\ }\href@noop {} {\bibfield  {journal} {\bibinfo
  {journal} {Prog. Solid St. Chem.}\ }\textbf {\bibinfo {volume} {15}},\
  \bibinfo {pages} {55} (\bibinfo {year} {1983})}\BibitemShut {NoStop}%
\bibitem [{\citenamefont
  {Rodr\'{i}guez-Carvajal}(1993)}]{Rodriguez-Carvajal1993}%
  \BibitemOpen
  \bibfield  {author} {\bibinfo {author} {\bibfnamefont {J.}~\bibnamefont
  {Rodr\'{i}guez-Carvajal}},\ }\href@noop {} {\bibfield  {journal} {\bibinfo
  {journal} {Phys. B: Condens. Matter.}\ }\textbf {\bibinfo {volume} {192}},\
  \bibinfo {pages} {55} (\bibinfo {year} {1993})}\BibitemShut {NoStop}%
\bibitem [{\citenamefont {Sala}\ \emph {et~al.}(2014)\citenamefont {Sala},
  \citenamefont {Gutmann}, \citenamefont {Prabhakaran}, \citenamefont
  {Pomaranski}, \citenamefont {Mitchelitis}, \citenamefont {Kycia},
  \citenamefont {Porter}, \citenamefont {Castelnovo},\ and\ \citenamefont
  {Goff}}]{Sala2014}%
  \BibitemOpen
  \bibfield  {author} {\bibinfo {author} {\bibfnamefont {G.}~\bibnamefont
  {Sala}}, \bibinfo {author} {\bibfnamefont {M.~J.}\ \bibnamefont {Gutmann}},
  \bibinfo {author} {\bibfnamefont {D.}~\bibnamefont {Prabhakaran}}, \bibinfo
  {author} {\bibfnamefont {D.}~\bibnamefont {Pomaranski}}, \bibinfo {author}
  {\bibfnamefont {C.}~\bibnamefont {Mitchelitis}}, \bibinfo {author}
  {\bibfnamefont {J.~B.}\ \bibnamefont {Kycia}}, \bibinfo {author}
  {\bibfnamefont {D.~G.}\ \bibnamefont {Porter}}, \bibinfo {author}
  {\bibfnamefont {C.}~\bibnamefont {Castelnovo}}, \ and\ \bibinfo {author}
  {\bibfnamefont {J.~P.}\ \bibnamefont {Goff}},\ }\href@noop {} {\bibfield
  {journal} {\bibinfo  {journal} {Nature Mater.}\ }\textbf {\bibinfo {volume}
  {13}},\ \bibinfo {pages} {488} (\bibinfo {year} {2014})}\BibitemShut
  {NoStop}%
\bibitem [{Ref()}]{Refinementdata}%
  \BibitemOpen
  \href@noop {} {}\bibinfo {note} {See Supplemental Material at [URL will be
  inserted by publisher] for further details of the structural
  refinement.}\BibitemShut {Stop}%
\bibitem [{\citenamefont {{Ciomaga Hatnean}}\ \emph
  {et~al.}(2014{\natexlab{b}})\citenamefont {{Ciomaga Hatnean}}, \citenamefont
  {Lees},\ and\ \citenamefont {Balakrishnan}}]{Hatnean2014_3}%
  \BibitemOpen
  \bibfield  {author} {\bibinfo {author} {\bibfnamefont {M.}~\bibnamefont
  {{Ciomaga Hatnean}}}, \bibinfo {author} {\bibfnamefont {M.~R.}\ \bibnamefont
  {Lees}}, \ and\ \bibinfo {author} {\bibfnamefont {G.}~\bibnamefont
  {Balakrishnan}},\ }\href@noop {} {} (\bibinfo {year} {2014}{\natexlab{b}}),\
  \bibinfo {note} {(unpublished)}\BibitemShut {NoStop}%
\bibitem [{\citenamefont {Hallas}\ \emph {et~al.}(2015)\citenamefont {Hallas},
  \citenamefont {Arevalo-Lopez}, \citenamefont {Sharma}, \citenamefont
  {Munsie}, \citenamefont {Attfield}, \citenamefont {Wiebe},\ and\
  \citenamefont {Luke}}]{Hallas15}%
  \BibitemOpen
  \bibfield  {author} {\bibinfo {author} {\bibfnamefont {A.~M.}\ \bibnamefont
  {Hallas}}, \bibinfo {author} {\bibfnamefont {A.~M.}\ \bibnamefont
  {Arevalo-Lopez}}, \bibinfo {author} {\bibfnamefont {A.~Z.}\ \bibnamefont
  {Sharma}}, \bibinfo {author} {\bibfnamefont {T.}~\bibnamefont {Munsie}},
  \bibinfo {author} {\bibfnamefont {J.~P.}\ \bibnamefont {Attfield}}, \bibinfo
  {author} {\bibfnamefont {C.~R.}\ \bibnamefont {Wiebe}}, \ and\ \bibinfo
  {author} {\bibfnamefont {G.~M.}\ \bibnamefont {Luke}},\ }\href@noop {}
  {\bibfield  {journal} {\bibinfo  {journal} {Phys. Rev. B}\ }\textbf {\bibinfo
  {volume} {91}},\ \bibinfo {pages} {104417} (\bibinfo {year}
  {2015})}\BibitemShut {NoStop}%
\bibitem [{\citenamefont {{Bondah-Jagalu}}\ and\ \citenamefont
  {Bramwell}(2001)}]{BondahJagalu01}%
  \BibitemOpen
  \bibfield  {author} {\bibinfo {author} {\bibfnamefont {V.}~\bibnamefont
  {{Bondah-Jagalu}}}\ and\ \bibinfo {author} {\bibfnamefont {S.~T.}\
  \bibnamefont {Bramwell}},\ }\href@noop {} {\bibfield  {journal} {\bibinfo
  {journal} {Can. J. Phys.}\ }\textbf {\bibinfo {volume} {79}},\ \bibinfo
  {pages} {1381} (\bibinfo {year} {2001})}\BibitemShut {NoStop}%
\bibitem [{\citenamefont {Matsuhira}\ \emph {et~al.}(2002)\citenamefont
  {Matsuhira}, \citenamefont {Hinatsu}, \citenamefont {Tenya}, \citenamefont
  {Amitsuka},\ and\ \citenamefont {Sakakibara}}]{Matsuhira2002}%
  \BibitemOpen
  \bibfield  {author} {\bibinfo {author} {\bibfnamefont {K.}~\bibnamefont
  {Matsuhira}}, \bibinfo {author} {\bibfnamefont {Y.}~\bibnamefont {Hinatsu}},
  \bibinfo {author} {\bibfnamefont {K.}~\bibnamefont {Tenya}}, \bibinfo
  {author} {\bibfnamefont {H.}~\bibnamefont {Amitsuka}}, \ and\ \bibinfo
  {author} {\bibfnamefont {T.}~\bibnamefont {Sakakibara}},\ }\href@noop {}
  {\bibfield  {journal} {\bibinfo  {journal} {J. Phys. Soc. Japan}\ }\textbf
  {\bibinfo {volume} {71}},\ \bibinfo {pages} {1576} (\bibinfo {year}
  {2002})}\BibitemShut {NoStop}%
\bibitem [{\citenamefont {Fukazawa}\ \emph {et~al.}(2002)\citenamefont
  {Fukazawa}, \citenamefont {Melko}, \citenamefont {Higashinaka}, \citenamefont
  {Maeno},\ and\ \citenamefont {Gingras}}]{Fukazawa2002}%
  \BibitemOpen
  \bibfield  {author} {\bibinfo {author} {\bibfnamefont {H.}~\bibnamefont
  {Fukazawa}}, \bibinfo {author} {\bibfnamefont {R.~G.}\ \bibnamefont {Melko}},
  \bibinfo {author} {\bibfnamefont {R.}~\bibnamefont {Higashinaka}}, \bibinfo
  {author} {\bibfnamefont {Y.}~\bibnamefont {Maeno}}, \ and\ \bibinfo {author}
  {\bibfnamefont {M.~J.~P.}\ \bibnamefont {Gingras}},\ }\href@noop {}
  {\bibfield  {journal} {\bibinfo  {journal} {Phys. Rev. B}\ }\textbf {\bibinfo
  {volume} {65}},\ \bibinfo {pages} {054410} (\bibinfo {year}
  {2002})}\BibitemShut {NoStop}%
\bibitem [{\citenamefont {Petrenko}\ \emph {et~al.}(2003)\citenamefont
  {Petrenko}, \citenamefont {Lees},\ and\ \citenamefont
  {Balakrishnan}}]{Petrenko2003}%
  \BibitemOpen
  \bibfield  {author} {\bibinfo {author} {\bibfnamefont {O.~A.}\ \bibnamefont
  {Petrenko}}, \bibinfo {author} {\bibfnamefont {M.~R.}\ \bibnamefont {Lees}},
  \ and\ \bibinfo {author} {\bibfnamefont {G.}~\bibnamefont {Balakrishnan}},\
  }\href@noop {} {\bibfield  {journal} {\bibinfo  {journal} {Phys. Rev. B}\
  }\textbf {\bibinfo {volume} {68}},\ \bibinfo {pages} {012406} (\bibinfo
  {year} {2003})}\BibitemShut {NoStop}%
\bibitem [{\citenamefont {Lutique}\ \emph {et~al.}(2003)\citenamefont
  {Lutique}, \citenamefont {Javorsk\'{y}}, \citenamefont {Konings},
  \citenamefont {van Genderen}, \citenamefont {van Miltenburg},\ and\
  \citenamefont {Wastin}}]{Lutique2003_2}%
  \BibitemOpen
  \bibfield  {author} {\bibinfo {author} {\bibfnamefont {S.}~\bibnamefont
  {Lutique}}, \bibinfo {author} {\bibfnamefont {P.}~\bibnamefont
  {Javorsk\'{y}}}, \bibinfo {author} {\bibfnamefont {R.~J.~M.}\ \bibnamefont
  {Konings}}, \bibinfo {author} {\bibfnamefont {A.~C.~G.}\ \bibnamefont {van
  Genderen}}, \bibinfo {author} {\bibfnamefont {J.~C.}\ \bibnamefont {van
  Miltenburg}}, \ and\ \bibinfo {author} {\bibfnamefont {F.}~\bibnamefont
  {Wastin}},\ }\href@noop {} {\bibfield  {journal} {\bibinfo  {journal} {J.
  Chem. Thermodynamics}\ }\textbf {\bibinfo {volume} {35}},\ \bibinfo {pages}
  {955} (\bibinfo {year} {2003})}\BibitemShut {NoStop}%
\bibitem [{\citenamefont {Crosswhite}\ and\ \citenamefont
  {Crosswhite}(1984)}]{Crosswhite1984}%
  \BibitemOpen
  \bibfield  {author} {\bibinfo {author} {\bibfnamefont {H.~M.}\ \bibnamefont
  {Crosswhite}}\ and\ \bibinfo {author} {\bibfnamefont {H.}~\bibnamefont
  {Crosswhite}},\ }\href@noop {} {\bibfield  {journal} {\bibinfo  {journal} {J.
  Opt. Soc. Am. B}\ }\textbf {\bibinfo {volume} {1}},\ \bibinfo {pages} {246}
  (\bibinfo {year} {1984})}\BibitemShut {NoStop}%
\bibitem [{\citenamefont {Malkin}\ \emph {et~al.}(2010)\citenamefont {Malkin},
  \citenamefont {Lummen}, \citenamefont {van Loosdrecht}, \citenamefont
  {Dhalenne},\ and\ \citenamefont {Zakirov}}]{Malkin2010}%
  \BibitemOpen
  \bibfield  {author} {\bibinfo {author} {\bibfnamefont {B.~Z.}\ \bibnamefont
  {Malkin}}, \bibinfo {author} {\bibfnamefont {T.~T.~A.}\ \bibnamefont
  {Lummen}}, \bibinfo {author} {\bibfnamefont {P.~H.~M.}\ \bibnamefont {van
  Loosdrecht}}, \bibinfo {author} {\bibfnamefont {G.}~\bibnamefont {Dhalenne}},
  \ and\ \bibinfo {author} {\bibfnamefont {A.~R.}\ \bibnamefont {Zakirov}},\
  }\href@noop {} {\bibfield  {journal} {\bibinfo  {journal} {J. Phys.: Condens.
  Matter.}\ }\textbf {\bibinfo {volume} {22}},\ \bibinfo {pages} {276003}
  (\bibinfo {year} {2010})}\BibitemShut {NoStop}%
\bibitem [{\citenamefont {Petit}\ \emph {et~al.}(2014)\citenamefont {Petit},
  \citenamefont {Robert}, \citenamefont {Guitteny}, \citenamefont {Bonville},
  \citenamefont {Decorse}, \citenamefont {Ollivier}, \citenamefont {Mutka},
  \citenamefont {Gingras},\ and\ \citenamefont {Mirebeau}}]{Petit2014}%
  \BibitemOpen
  \bibfield  {author} {\bibinfo {author} {\bibfnamefont {S.}~\bibnamefont
  {Petit}}, \bibinfo {author} {\bibfnamefont {J.}~\bibnamefont {Robert}},
  \bibinfo {author} {\bibfnamefont {S.}~\bibnamefont {Guitteny}}, \bibinfo
  {author} {\bibfnamefont {P.}~\bibnamefont {Bonville}}, \bibinfo {author}
  {\bibfnamefont {C.}~\bibnamefont {Decorse}}, \bibinfo {author} {\bibfnamefont
  {J.}~\bibnamefont {Ollivier}}, \bibinfo {author} {\bibfnamefont
  {H.}~\bibnamefont {Mutka}}, \bibinfo {author} {\bibfnamefont {M.~J.~P.}\
  \bibnamefont {Gingras}}, \ and\ \bibinfo {author} {\bibfnamefont
  {I.}~\bibnamefont {Mirebeau}},\ }\href@noop {} {\bibfield  {journal}
  {\bibinfo  {journal} {Phys. Rev. B}\ }\textbf {\bibinfo {volume} {90}},\
  \bibinfo {pages} {060410} (\bibinfo {year} {2014})}\BibitemShut {NoStop}%
\bibitem [{\citenamefont {Popova}\ \emph {et~al.}(2007)\citenamefont {Popova},
  \citenamefont {Chukalina}, \citenamefont {Stanislavchuk}, \citenamefont
  {Malkin}, \citenamefont {Zakirov}, \citenamefont {Antic-Fidancev},
  \citenamefont {Popova}, \citenamefont {Bezmaternykh},\ and\ \citenamefont
  {Temerov}}]{Popova2007}%
  \BibitemOpen
  \bibfield  {author} {\bibinfo {author} {\bibfnamefont {M.~N.}\ \bibnamefont
  {Popova}}, \bibinfo {author} {\bibfnamefont {E.~P.}\ \bibnamefont
  {Chukalina}}, \bibinfo {author} {\bibfnamefont {T.~N.}\ \bibnamefont
  {Stanislavchuk}}, \bibinfo {author} {\bibfnamefont {B.~Z.}\ \bibnamefont
  {Malkin}}, \bibinfo {author} {\bibfnamefont {A.~R.}\ \bibnamefont {Zakirov}},
  \bibinfo {author} {\bibfnamefont {E.}~\bibnamefont {Antic-Fidancev}},
  \bibinfo {author} {\bibfnamefont {E.~A.}\ \bibnamefont {Popova}}, \bibinfo
  {author} {\bibfnamefont {L.~N.}\ \bibnamefont {Bezmaternykh}}, \ and\
  \bibinfo {author} {\bibfnamefont {V.~L.}\ \bibnamefont {Temerov}},\
  }\href@noop {} {\bibfield  {journal} {\bibinfo  {journal} {Phys. Rev. B}\
  }\textbf {\bibinfo {volume} {75}},\ \bibinfo {pages} {224435} (\bibinfo
  {year} {2007})}\BibitemShut {NoStop}%
\bibitem [{\citenamefont {Klekovkina}\ and\ \citenamefont
  {Malkin}(2014)}]{Klekovkina2014}%
  \BibitemOpen
  \bibfield  {author} {\bibinfo {author} {\bibfnamefont {V.~V.}\ \bibnamefont
  {Klekovkina}}\ and\ \bibinfo {author} {\bibfnamefont {B.~Z.}\ \bibnamefont
  {Malkin}},\ }\href@noop {} {\bibfield  {journal} {\bibinfo  {journal} {Opt.
  Spectrosc.}\ }\textbf {\bibinfo {volume} {116}},\ \bibinfo {pages} {849}
  (\bibinfo {year} {2014})}\BibitemShut {NoStop}%
\bibitem [{\citenamefont {Malkin}\ \emph {et~al.}(2004)\citenamefont {Malkin},
  \citenamefont {Zakirov}, \citenamefont {N.Popova}, \citenamefont {Klimin},
  \citenamefont {Chukalina}, \citenamefont {E.Antic-Fidancev}, \citenamefont
  {Goldner}, \citenamefont {Aschehoug},\ and\ \citenamefont
  {Dhalenne}}]{Malkin2004}%
  \BibitemOpen
  \bibfield  {author} {\bibinfo {author} {\bibfnamefont {B.~Z.}\ \bibnamefont
  {Malkin}}, \bibinfo {author} {\bibfnamefont {A.~R.}\ \bibnamefont {Zakirov}},
  \bibinfo {author} {\bibfnamefont {M.}~\bibnamefont {N.Popova}}, \bibinfo
  {author} {\bibfnamefont {S.~A.}\ \bibnamefont {Klimin}}, \bibinfo {author}
  {\bibfnamefont {E.~P.}\ \bibnamefont {Chukalina}}, \bibinfo {author}
  {\bibnamefont {E.Antic-Fidancev}}, \bibinfo {author} {\bibfnamefont {{\relax
  Ph}.}~\bibnamefont {Goldner}}, \bibinfo {author} {\bibfnamefont
  {P.}~\bibnamefont {Aschehoug}}, \ and\ \bibinfo {author} {\bibnamefont
  {Dhalenne}},\ }\href@noop {} {\bibfield  {journal} {\bibinfo  {journal}
  {Phys. Rev. B}\ }\textbf {\bibinfo {volume} {70}},\ \bibinfo {pages} {075112}
  (\bibinfo {year} {2004})}\BibitemShut {NoStop}%
\bibitem [{\citenamefont {Sedmidubsk\'{y}}\ \emph {et~al.}(2005)\citenamefont
  {Sedmidubsk\'{y}}, \citenamefont {Bene\u{s}},\ and\ \citenamefont
  {Konings}}]{Sedmidubsky2005}%
  \BibitemOpen
  \bibfield  {author} {\bibinfo {author} {\bibfnamefont {D.}~\bibnamefont
  {Sedmidubsk\'{y}}}, \bibinfo {author} {\bibfnamefont {O.}~\bibnamefont
  {Bene\u{s}}}, \ and\ \bibinfo {author} {\bibfnamefont {R.~J.~M.}\
  \bibnamefont {Konings}},\ }\href@noop {} {\bibfield  {journal} {\bibinfo
  {journal} {J. Chem. Thermodynamics}\ }\textbf {\bibinfo {volume} {37}},\
  \bibinfo {pages} {1098} (\bibinfo {year} {2005})}\BibitemShut {NoStop}%
\bibitem [{\citenamefont {Lan}\ \emph {et~al.}()\citenamefont {Lan},
  \citenamefont {Ouyang},\ and\ \citenamefont {Song}}]{Lan15}%
  \BibitemOpen
  \bibfield  {author} {\bibinfo {author} {\bibfnamefont {G.}~\bibnamefont
  {Lan}}, \bibinfo {author} {\bibfnamefont {B.}~\bibnamefont {Ouyang}}, \ and\
  \bibinfo {author} {\bibfnamefont {J.}~\bibnamefont {Song}},\ }\href@noop {}
  {}\bibinfo {note} {{a}rXiv 1503.03875}\BibitemShut {NoStop}%
\bibitem [{\citenamefont {Lutique}\ \emph {et~al.}(2004)\citenamefont
  {Lutique}, \citenamefont {Javorsk\'{y}}, \citenamefont {Konings},
  \citenamefont {Krupa}, \citenamefont {van Genderen}, \citenamefont {van
  Miltenburg},\ and\ \citenamefont {Wastin}}]{Lutique2004}%
  \BibitemOpen
  \bibfield  {author} {\bibinfo {author} {\bibfnamefont {S.}~\bibnamefont
  {Lutique}}, \bibinfo {author} {\bibfnamefont {P.}~\bibnamefont
  {Javorsk\'{y}}}, \bibinfo {author} {\bibfnamefont {R.~J.~M.}\ \bibnamefont
  {Konings}}, \bibinfo {author} {\bibfnamefont {J.-C.}\ \bibnamefont {Krupa}},
  \bibinfo {author} {\bibfnamefont {A.~C.~G.}\ \bibnamefont {van Genderen}},
  \bibinfo {author} {\bibfnamefont {J.~C.}\ \bibnamefont {van Miltenburg}}, \
  and\ \bibinfo {author} {\bibfnamefont {F.}~\bibnamefont {Wastin}},\
  }\href@noop {} {\bibfield  {journal} {\bibinfo  {journal} {J. Chem.
  Thermodynamics}\ }\textbf {\bibinfo {volume} {36}},\ \bibinfo {pages} {609}
  (\bibinfo {year} {2004})}\BibitemShut {NoStop}%
\bibitem [{\citenamefont {Princep}\ \emph {et~al.}(2013)\citenamefont
  {Princep}, \citenamefont {Prabhakaran}, \citenamefont {Boothroyd},\ and\
  \citenamefont {Adroja}}]{Princep2013}%
  \BibitemOpen
  \bibfield  {author} {\bibinfo {author} {\bibfnamefont {A.~J.}\ \bibnamefont
  {Princep}}, \bibinfo {author} {\bibfnamefont {D.}~\bibnamefont
  {Prabhakaran}}, \bibinfo {author} {\bibfnamefont {A.~T.}\ \bibnamefont
  {Boothroyd}}, \ and\ \bibinfo {author} {\bibfnamefont {D.~T.}\ \bibnamefont
  {Adroja}},\ }\href@noop {} {\bibfield  {journal} {\bibinfo  {journal} {Phys.
  Rev. B}\ }\textbf {\bibinfo {volume} {88}},\ \bibinfo {pages} {104421}
  (\bibinfo {year} {2013})}\BibitemShut {NoStop}%
\bibitem [{\citenamefont {Watahiki}\ \emph {et~al.}(2011)\citenamefont
  {Watahiki}, \citenamefont {Tomiyasu}, \citenamefont {Matsuhira},
  \citenamefont {Iwasa}, \citenamefont {Yokoyama}, \citenamefont {Takagi},
  \citenamefont {Wakeshima},\ and\ \citenamefont {Hinatsu}}]{Watahiki2011}%
  \BibitemOpen
  \bibfield  {author} {\bibinfo {author} {\bibfnamefont {M.}~\bibnamefont
  {Watahiki}}, \bibinfo {author} {\bibfnamefont {K.}~\bibnamefont {Tomiyasu}},
  \bibinfo {author} {\bibfnamefont {K.}~\bibnamefont {Matsuhira}}, \bibinfo
  {author} {\bibfnamefont {K.}~\bibnamefont {Iwasa}}, \bibinfo {author}
  {\bibfnamefont {M.}~\bibnamefont {Yokoyama}}, \bibinfo {author}
  {\bibfnamefont {S.}~\bibnamefont {Takagi}}, \bibinfo {author} {\bibfnamefont
  {M.}~\bibnamefont {Wakeshima}}, \ and\ \bibinfo {author} {\bibfnamefont
  {Y.}~\bibnamefont {Hinatsu}},\ }\href@noop {} {\bibfield  {journal} {\bibinfo
   {journal} {J. Phys.: Conf. Ser.}\ }\textbf {\bibinfo {volume} {320}},\
  \bibinfo {pages} {012080} (\bibinfo {year} {2011})}\BibitemShut {NoStop}%
\bibitem [{\citenamefont {Huang}\ \emph {et~al.}(2014)\citenamefont {Huang},
  \citenamefont {Chen},\ and\ \citenamefont {Hermele}}]{Huang2014}%
  \BibitemOpen
  \bibfield  {author} {\bibinfo {author} {\bibfnamefont {Y.-P.}\ \bibnamefont
  {Huang}}, \bibinfo {author} {\bibfnamefont {G.}~\bibnamefont {Chen}}, \ and\
  \bibinfo {author} {\bibfnamefont {M.}~\bibnamefont {Hermele}},\ }\href@noop
  {} {\bibfield  {journal} {\bibinfo  {journal} {Phys. Rev. Lett.}\ }\textbf
  {\bibinfo {volume} {112}},\ \bibinfo {pages} {167203} (\bibinfo {year}
  {2014})}\BibitemShut {NoStop}%
\bibitem [{\citenamefont {Yasui}\ \emph {et~al.}(2003)\citenamefont {Yasui},
  \citenamefont {Iikubo}, \citenamefont {Harashina}, \citenamefont {Kageyama},
  \citenamefont {Ito}, \citenamefont {Sato},\ and\ \citenamefont
  {Kakurai}}]{Yasui2003}%
  \BibitemOpen
  \bibfield  {author} {\bibinfo {author} {\bibfnamefont {Y.}~\bibnamefont
  {Yasui}}, \bibinfo {author} {\bibfnamefont {S.}~\bibnamefont {Iikubo}},
  \bibinfo {author} {\bibfnamefont {H.}~\bibnamefont {Harashina}}, \bibinfo
  {author} {\bibfnamefont {T.}~\bibnamefont {Kageyama}}, \bibinfo {author}
  {\bibfnamefont {M.}~\bibnamefont {Ito}}, \bibinfo {author} {\bibfnamefont
  {M.}~\bibnamefont {Sato}}, \ and\ \bibinfo {author} {\bibfnamefont
  {K.}~\bibnamefont {Kakurai}},\ }\href@noop {} {\bibfield  {journal} {\bibinfo
   {journal} {J. Phys. Soc. Jpn.}\ }\textbf {\bibinfo {volume} {72}},\ \bibinfo
  {pages} {865} (\bibinfo {year} {2003})}\BibitemShut {NoStop}%
\bibitem [{\citenamefont {Onose}\ \emph {et~al.}(2004)\citenamefont {Onose},
  \citenamefont {Taguchi}, \citenamefont {Ito},\ and\ \citenamefont
  {Tokura}}]{Onose2004}%
  \BibitemOpen
  \bibfield  {author} {\bibinfo {author} {\bibfnamefont {Y.}~\bibnamefont
  {Onose}}, \bibinfo {author} {\bibfnamefont {Y.}~\bibnamefont {Taguchi}},
  \bibinfo {author} {\bibfnamefont {T.}~\bibnamefont {Ito}}, \ and\ \bibinfo
  {author} {\bibfnamefont {Y.}~\bibnamefont {Tokura}},\ }\href@noop {}
  {\bibfield  {journal} {\bibinfo  {journal} {Phys. Rev. B}\ }\textbf {\bibinfo
  {volume} {70}},\ \bibinfo {pages} {060401(R)} (\bibinfo {year}
  {2004})}\BibitemShut {NoStop}%
\bibitem [{\citenamefont {Levy}(1964)}]{Levy1964}%
  \BibitemOpen
  \bibfield  {author} {\bibinfo {author} {\bibfnamefont {P.~M.}\ \bibnamefont
  {Levy}},\ }\href@noop {} {\bibfield  {journal} {\bibinfo  {journal} {Phys.
  Rev.}\ }\textbf {\bibinfo {volume} {135}},\ \bibinfo {pages} {A155} (\bibinfo
  {year} {1964})}\BibitemShut {NoStop}%
\bibitem [{\citenamefont {Ewing}\ \emph {et~al.}(2004)\citenamefont {Ewing},
  \citenamefont {Weber},\ and\ \citenamefont {Lian}}]{Ewing2004}%
  \BibitemOpen
  \bibfield  {author} {\bibinfo {author} {\bibfnamefont {R.~C.}\ \bibnamefont
  {Ewing}}, \bibinfo {author} {\bibfnamefont {W.~J.}\ \bibnamefont {Weber}}, \
  and\ \bibinfo {author} {\bibfnamefont {J.}~\bibnamefont {Lian}},\ }\href@noop
  {} {\bibfield  {journal} {\bibinfo  {journal} {J. Appl. Phys.}\ }\textbf
  {\bibinfo {volume} {95}},\ \bibinfo {pages} {5949} (\bibinfo {year}
  {2004})}\BibitemShut {NoStop}%
\bibitem [{\citenamefont {Wu}\ \emph {et~al.}(2002)\citenamefont {Wu},
  \citenamefont {Wei}, \citenamefont {Padture}, \citenamefont {Klemens},
  \citenamefont {Gell}, \citenamefont {Garcia}, \citenamefont {Miranzo},\ and\
  \citenamefont {Osendi}}]{Wu2002}%
  \BibitemOpen
  \bibfield  {author} {\bibinfo {author} {\bibfnamefont {J.}~\bibnamefont
  {Wu}}, \bibinfo {author} {\bibfnamefont {X.}~\bibnamefont {Wei}}, \bibinfo
  {author} {\bibfnamefont {N.~P.}\ \bibnamefont {Padture}}, \bibinfo {author}
  {\bibfnamefont {P.~G.}\ \bibnamefont {Klemens}}, \bibinfo {author}
  {\bibfnamefont {M.}~\bibnamefont {Gell}}, \bibinfo {author} {\bibfnamefont
  {E.}~\bibnamefont {Garcia}}, \bibinfo {author} {\bibfnamefont
  {P.}~\bibnamefont {Miranzo}}, \ and\ \bibinfo {author} {\bibfnamefont
  {M.~I.}\ \bibnamefont {Osendi}},\ }\href@noop {} {\bibfield  {journal}
  {\bibinfo  {journal} {J. Am. Ceram. Soc.}\ }\textbf {\bibinfo {volume}
  {85}},\ \bibinfo {pages} {3031} (\bibinfo {year} {2002})}\BibitemShut
  {NoStop}%
\end{thebibliography}%
\end{document}